\documentclass[preprint,12pt]{elsarticle}

\usepackage{mathptmx} % 使用 Times 风格的字体
\usepackage{xcolor}
\usepackage{times}
\usepackage{epsfig}
\usepackage{epstopdf}
\usepackage{multicol}
\usepackage{array}
\usepackage{color}
\usepackage{colortbl}
\usepackage{bbding}
\usepackage[shortlabels]{enumitem}    % 编号扩展功能包
\usepackage{graphicx}    % 图片功能包
\usepackage{booktabs}    % 表格功能包
\usepackage{tablefootnote}
\usepackage{threeparttable}
\usepackage{multirow}    % 合并多行表格
\usepackage{amsmath}    % 公式功能包
\usepackage{amsthm}
\usepackage{amsfonts}
\usepackage{subcaption}
\usepackage{bm}
\usepackage{subeqnarray}
\usepackage{cases}
\newcounter{defcounter}
\usepackage{pdfpages}

\usepackage[utf8x]{inputenc}
\usepackage{lineno}
\usepackage{url}
\newcommand{\cut}[1]{}
\newcommand{\BibTeX}{\textrm{B \kern -.05em \textsc{i \kern -.025em b} \kern -.08em
T \kern -.1667em \lower .7ex \hbox{E} \kern -.125emX}}
\usepackage{verbatim}
\usepackage{indentfirst}
\usepackage[breaklinks=true,linkcolor=black,citecolor=black,bookmarks=false]{hyperref}
\hypersetup{
    colorlinks=true,
    linkcolor=black
}
\usepackage{textcomp}
\usepackage{mathtools}
\usepackage{wasysym}
\usepackage{tabularx}

\begin{document}

    \begin{frontmatter}



        \title{Multi-periodicity dependency Transformer based on spectrum offset for radio frequency fingerprint identification}

        \cortext[cor1]{Corresponding author}

        \author[buaa]{Jing Xiao}
        \ead{xiaoj@buaa.edu.cn}
        \affiliation[buaa]{organization={School of Electronics and Information Engineering, Beihang University},
            city={Beijing},
            postcode={100191},
            country={China}}

        \author[uas]{Wenrui Ding}
        \ead{ding@buaa.edu.cn}
        \affiliation[uas]{organization={Institute of Unmanned System, Beihang University},
            city={Beijing},
            postcode={100191},
            country={China}}

        \author[buaa]{Zeqi Shao}
        \ead{shaozeqi@buaa.edu.cn}

        \author[ncut]{Duona Zhang\corref{cor1}}
        \ead{zhangduona@ncut.edu.cn}
        \affiliation[ncut]{organization={School of Information Science and Technology, North China University of Technology},
            city={Beijing},
            postcode={100144},
            country={China}}

        \author[jskxy]{Yanan Ma\corref{cor1}}
        \ead{ynma2019@163.com}

        \author[uas]{Yufeng Wang}
        \ead{wyfeng@buaa.edu.cn}

        \author[jskxy]{Jian Wang}
        \ead{wangjian710108@126.com}

        \affiliation[jskxy]{organization={National Innovation Institute of Defense Technology, Academy of Military Sciences},
            city={Beijing},
            postcode={100071},
            country={China}}

        \begin{abstract}

            Radio Frequency Fingerprint Identification (RFFI) has emerged as a pivotal task for reliable device authentication. Despite advancements in RFFI methods, background noise and intentional modulation features result in weak energy and subtle differences in the RFF features. These challenges diminish the capability of RFFI methods in feature representation, complicating the effective identification of device identities. This paper proposes a novel Multi-Periodicity Dependency Transformer (MPDFormer) to address these challenges. The MPDFormer employs a spectrum offset-based periodic embedding representation to augment the discrepency of intrinsic features. We delve into the intricacies of the periodicity-dependency attention mechanism, integrating both inter-period and intra-period attention mechanisms. This mechanism facilitates the extraction of both long and short-range periodicity-dependency features , accentuating the feature distinction whilst concurrently attenuating the perturbations caused by background noise and weak-periodicity features. Empirical results demonstrate MPDFormer's superiority over established baseline methods, achieving a 0.07s inference time on NVIDIA Jetson Orin NX.

        \end{abstract}


    \end{frontmatter}

    \section{Introduction}

    The Internet of Things (IoT) has witnessed a meteoric rise in recent years, driven by its significant impact in various fields like smart healthcare, manufacturing, and urban infrastructure, etc~\cite{lu2014ConnectedVehiclesSolutions, younan2020ChallengesRecommendedTechnologies}.
    The proliferation of wireless devices from these fields~\cite{alavi2018InternetThingsenabledSmart}, however, has led to an increase in security risks, particularly in identity verification and data transmission.
    Amidst the evolving domain of wireless communication, traditional cryptographic methods face escalating threats, including sophisticated decryption techniques, key leakage, and systemic breaches~\cite{zhang2014SybilAttacksTheir, he2015AnalysisRFIDAuthentication}.
    These urgent needs accentuate the imperative for robust, alternative security paradigms.
    Radio Frequency Fingerprint Identification (RFFI), an emergent authentication mechanism, has surfaced as a promising countermeasure to these security threats.
    RFFI capitalizes on the hardware impairments intrinsic to analog components, a byproduct of manufacturing variations, to uniquely identify wireless devices~\cite{danev2012PhysicalLayerIdentificationWireless}.
    Similar to the biometric fingerprinting technique, RFFI technique exploits the intrinsic and unique features of Radio Frequency (RF) emissions to secure data transactions.

    Predominantly, the RFFI methods can be categorized into two groups: transient-based and steady state-based methods~\cite{jagannath2022ComprehensiveSurveyRadio}.
    Transient-based RFFI hinges on exploiting ephemeral state transitions within transmitters.
    The primary obstacle is the accurate start-point of signal amidst ambient noise, necessitating sophisticated apparatus such as high-speed oscilloscopes for capturing transient phenomena on the order of nanoseconds~\cite{yu2019RobustRFFingerprinting}.
    This requisite impinges upon the practicability of transient-based RFFI methods, constricting the acquisition of substantial transient signals.
    Conversely, steady state-based RFFI offers a more pragmatic approach, availing itself of stable operational states conducive to robust device characterization~\cite{tian2019NewSecurityMechanisms}.
    These methods concentrate on distilling unique signatures from the stable phase of device operations, wherein hardware inconsistencies become discernible and instrumental in device identification.
    In recent years, Deep Learning (DL) methods employed within RFFI frameworks have been paid more attention for their effectiveness in extracting intricate nonlinear features intrinsic to RFFs, outperforming conventional RFFI methods~\cite{sankhe2019ORACLEOptimizedRadio}.

    Despite significant advancements in RFFI methods, several challenges still persist.
    On one hand, intrinsic Radio Frequency Fingerprint (RFF) features are vulnerable to interference from background noise, nonlinear channel effects, and intentional modulation, leading to nuanced discrepancies in the RFF signatures of various wireless devices.
    These subtle variations pose a considerable obstacle to the effective extraction, differentiation, and identification of RFF features using current RFFI methods.
    On the other hand, RFF features stem from a myriad of hardware impairments.
    While the collective impact of RFF features remains stable, the inherent randomness of hardware components in individual devices renders the differentiation of RFF features challenging to discern in the time domain.
    Consequently, these issues present considerable hurdles for the application of RFFI methods in real-world scenarios.

    To address the aforementioned challenges, this paper proposes an innovative Multi-Periodicity Dependency Transformer (MPDFormer) architecture based on deep learning.
    Recognizing the complexity of discerning RFF feature differences in the time domain, MPDFormer focuses on frequency-domain differentiation based on spectrum offset.
    Building upon these frequency-domain differentiation, the architecture achieves multiple Periodic Embedding Representation (PER) in the time domain.
    Although the differences in RFF features are subtle, fingerprint characteristics repetitively emerge in different segments of the time-series signal, exhibiting periodic traits.
    Therefore, we propose a periodicity-dependency attention mechanism, composing of inter-period and intra-period attention mechanisms, to enhance the RFF features across multiple specific periods while concurrently suppressing background noise and features of weak-periodicity or non-specified periodicity.
    Finally, MPDFormer employs an multi-periodicity feature adaptive-fusion module for selectively enhancing the varying-dimension RFF features at the specified periodicity, thereby empowering the representation capability of the proposed RFFI method.
    With the employment on a edge computing platform, the MPDFormer model achieves impressive inference times of under 0.1 seconds on the NVIDIA Jetson Orin NX.
    Our key contributions are summarized as follows:

    \begin{itemize}
    [label=$\bullet$]
        \item Our method leverages spectrum offset to prioritize periods with maximal differences in RFF features for embedding representation, integrating domain knowledge-based time-series analysis to enhance feature extraction without exhaustive enumeration in Transformer architecture networks.

        \item The MPDFormer integrates both inter-period and intra-period attention mechanisms, capturing long and short-range periodic dependencies. This enhances RFF feature distinctiveness across devices, reducing the impact of background noise and weak periodicity.

        \item RF signals from 32 ZigBee devices were collected for the ZigBeeRFF2023 dataset, which will soon be available. The dataset includes two subsets showing consistent and random data patterns, facilitating the analysis of device identification performance under varying conditions.

        \item The proposed method has been deployed on the NVIDIA Jetson Orin NX for practical edge computing, demonstrating high real-time performance with inference times of millisecond level, adequate for most practical wireless communication scenarios.

        \item Compared with existing methods, MPDFormer outperforms in low SNR scenarios using ZigBee signal datasets, achieving better intra-class compactness and inter-class dispersion, and offering robustness against various disturbances.
    \end{itemize}

    The following sections are structured as follows:
    Section~\ref{sec:related-work} presents the related work.
    Section~\ref{sec:proposed-methodology} delves into the proposed method.
    Section~\ref{sec:signal-acquisition-and-collection} describes the acquisition and collection of the RF signals.
    Section~\ref{sec:experiment-and-results} discusses the experimental results.
    Finally, Section VI concludes the paper, offering insights into future research directions.

    \hspace{-1pt}
    \begin{figure} [t]
        \begin{center}
            \includegraphics[width=0.8\textwidth]{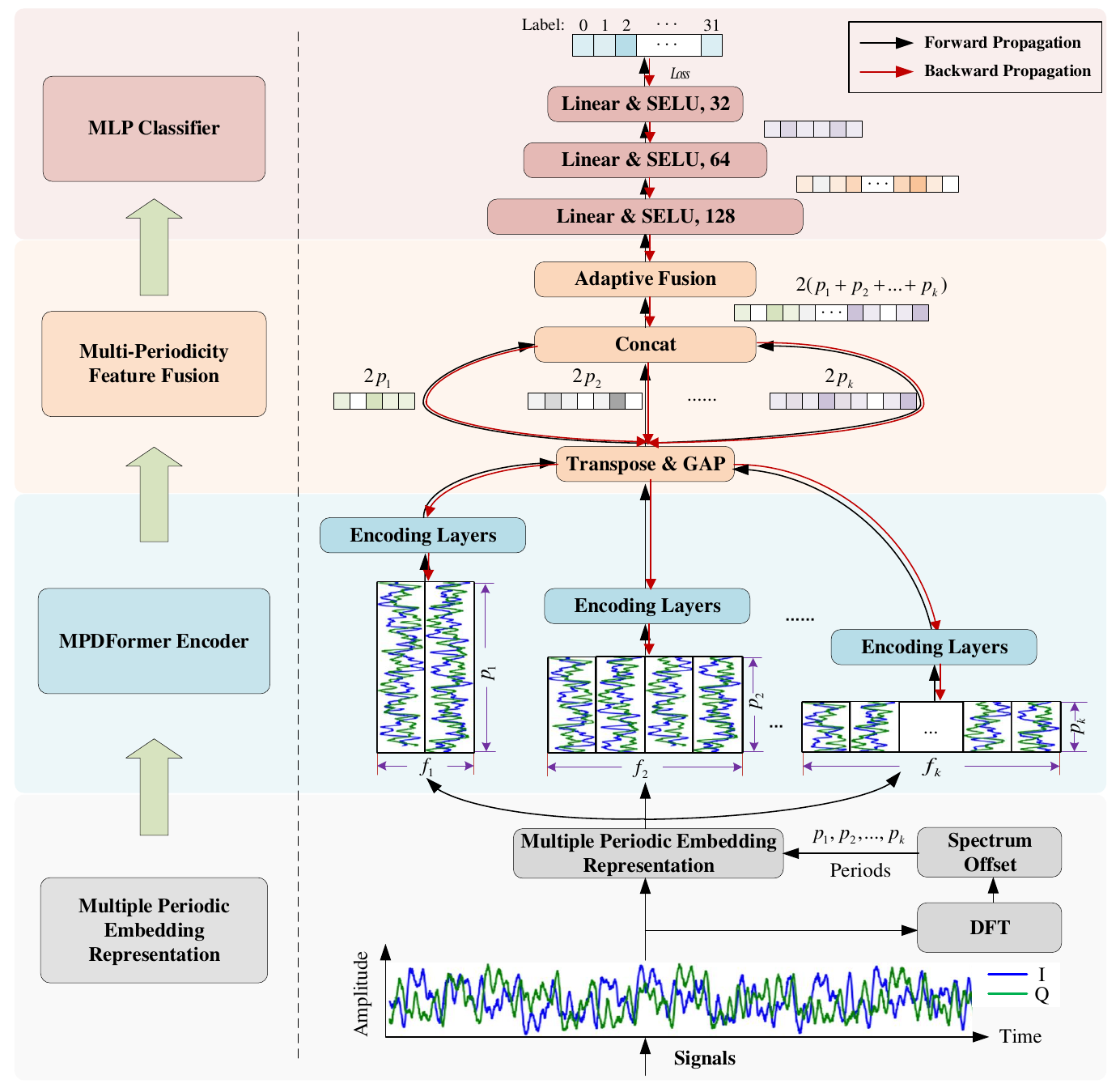}
        \end{center}
        \caption{Network architecture of the proposed MPDFormer.}
        \label{fig:total_network_architecture}
    \end{figure}

    \section{Related work}\label{sec:related-work}

    RFFI constitutes a pivotal innovation in leveraging distinctive RF emitter characteristics for precision-based device identification.
    This domain has evolved from its nascent stages, which predominantly relied on expert feature extraction and transient-based RFFI techniques, to embrace more sophisticated steady state-based RFFI methods.
    Our discourse concentrates on two primary categories within this field: transient-based and steady state-based methods~\cite{soltanieh2020ReviewRadioFrequency}.

    \subsection{Transient-based methods}

    Transient-based methods exploit the unique characteristics manifested during the transient phase of a signal, specifically during operation states, such as powering on or off.
    These transient phases give rise to the rise-time and fall-time signal states, unveiling distinctive traits influenced by the hardware impairments.
    By meticulously analyzing the transient response, these methods capitalize on the complex patterns and variations within transient features to distill discriminative information.
    Therefore, the transient-based RFFI method is employed initially for precise start point of signal, followed by the extraction of RFF features from RF transient signals, thus enabling the identification of wireless devices.

    Significant progress has already been achieved in the detection methods for pinpointing the precise start point of a signal in previous studies.
    The start point detection of RF signal relies on methods such as threshold detection, Bayesian step change detection, and generalized likelihood ratio hypothesis testing.
    Shaw and Kinsner~\cite{shaw1997MultifractalModellingRadio} innovatively utilized variance fractal dimension trajectories and threshold setting to segregate RF signals from noisy channels, yielding satisfactory detection outcomes.
    Subsequent studies~\cite{higuchi1988ApproachIrregularTime, ureten1999DetectionRadioTransmitter, ureten2005BayesianDetectionWiFi} employed Bayesian step change detection for RF signal onset identification, eliminating the need for threshold setting.
    Selcuk et al~\cite{tascioglu2022SequentialTransientDetection} employed a generalized likelihood ratio algorithm approximation to detect transient signal onsets without any priori knowledge.

    Afterwards, transient-based methods leverage the nuances of transient signals to distill discriminative features for wireless device identification.
    The application of the Hilbert-Huang Transform has been instrumental in clarifying time-frequency distributions of transient-state signals, thereby enabling the extraction of RFF features by capturing thirteen distinct attributes~\cite{yuan2014SpecificEmitterIdentification}.
    In the time-frequency domain, the transient energy spectrum is derived using techniques such as the short-time Fourier transform~\cite{urrehman2012RFFingerprintExtraction, zhao2013WirelessLocalArea} or the Discrete Fourier Transform (DFT)~\cite{kose2019RFFingerprintingIoT}.
    Additionally, multifractal features intrinsic to instantaneous transient signal envelopes are extracted and utilized for device identification through the deployment of a support vector machine classifier~\cite{shi2015MultifractalSlopeFeature}.

    Nonetheless, certain methods~\cite{higuchi1988ApproachIrregularTime, shaw1997MultifractalModellingRadio, hall2003DetectionTransientRadio, ureten2005BayesianDetectionWiFi} demonstrate a strong sensitivity to noise interference during the detection of a signal's start point.
    This sensitivity introduces substantial challenges in accurate signal acquisition.
    Moreover, the task of discerning RF signals from ambient noise in a non-stationary state is notably arduous~\cite{zhao2013WirelessLocalArea}.
    Compounding these challenges are the inherent vulnerabilities of transient signals to multipath propagation, Doppler effect, and other non-linear distortions, collectively diminishing the efficacy of transient-based methods in practical, real-world scenarios.

    \subsection{Steady state-based methods}

    Conversely, steady state-based methods bypass the transient phase, focusing instead on the stable attributes exhibited during a signal's steady-state operation.
    By isolating and scrutinizing the continuous facets of the signal, these methods exploit inherent features such as amplitude, phase, and frequency characteristics~\cite{satija2019SpecificEmitterIdentification}.
    These features, stemming from the steady-state response, aid in differentiating various RF signals based on their consistent behaviors.
    Therefore, steady state-based techniques are particularly advantageous when dealing with signals that exhibit a relatively constant state over prolonged durations in wireless communication systems.

    Conventionally, steady state-based methods predominantly depend on manually extracted features, including statistical, time-domain, and transform-domain features.
    In terms of statistical features, Hiren~\cite{patel2015NonparametricFeatureGeneration} introduced non-parametric methods, encompassing metrics such as mean, median, mode, and linear model coefficient estimates, to more accurately depict the non-parametric distribution of ZigBee preamble signals.
    Furthermore, by integrating instantaneous features in the time domain, these RFF features have demonstrated enhanced efficacy compared to conventional RFF features reliant on instantaneous amplitude, phase, frequency, etc.
    This method holds promise for bolstering wireless network security.
    Nonetheless, it is imperative to note that methods contingent on features like instantaneous amplitude, phase, and envelope, as well as their statistical parameters in the time domain, are prone to noise interference.

    To mitigate the effect of noise interference, Randall W. et al.~\cite{klein2009ApplicationWaveletbasedRF} substantiated the efficacy of RFF features in the wavelet domain utilizing dual-tree complex wavelet transform for intra-manufacturer classification of identical model Cisco devices.
    This method offers augmented device distinguishability compared to traditional RFF features in the time domain, particularly under SNRs.
    Shen~\cite{shen2021RadioFrequencyFingerprint} undertook a comprehensive analysis within the impact of carrier frequency offset on system stability.
    The study evaluated diverse signal representations in time, frequency, and time-frequency domains, employing a hybrid classifier to refine classification accuracy by calibrating unreliable predictions.

    Recently, advancements in DL~\cite{elsisi2022EffectiveIoTbasedDeep, tran2023MachineLearningIoTbased, menagadevi2024MachineDeepLearning} have paved the way for the deployment of deep neural networks for the autonomous extraction of robust RFF features, showcasing considerable promise.
    Jian et al.~\cite{jian2021RadioFrequencyFingerprinting} pioneer the implementation of DL for RFF in edge devices, significantly enhancing the computational efficiency with minimal impact on accuracy and providing a foundational step towards the deployment of RFF on resource-constrained platforms.
    Peng et al.~\cite{peng2020DeepLearningBaseda} propose a DL-based RFF identification method utilizing differential constellation trace figure, which circumvents the need for signal synchronization and showcases a notable improvement in identification accuracy, thereby setting a new benchmark for IoT device authentication.
    The current realm of DL-based methods encounters a myriad of intricate challenges pertinent to the domain of RFF-based device identification.
    These challenges manifest in several forms, notably through intentional modulation~\cite{sun2022RadioFrequencyFingerprintExtraction}, the dynamics of the wireless communication channel~\cite{riyaz2018DeepLearningConvolutional, peng2019DesignHybridRF, zhou2021RobustRadioFrequencyFingerprint}, and the pervasive issue of noise interference~\cite{xing2018RadioFrequencyFingerprint, wu2022DSLNSecuringInternet}.

    As aforementioned above, transient-based RFFI methods are constrained by the necessity of precise start-point detection of signal, rendering them susceptible to noise interference and challenging for practical application.
    Moreover, current steady-state RFF methods, despite their commendable performance, are prone to being overwhelmed by intentionally modulated features, leading to the overfitting of non-RFF features in limited sample size and overlooking the intrinsic RFF features.
    Inspired by previous studies, we propose an novel RFF feature metric method based on spectrum offset in the frequency domain, which characterized with the multi-periodicity dependency signal components in the time domain.
    Subsequently, we utilize a periodicity-dependency attention mechanism based on Transformer architecture to mitigate the effects of intentional modulation features, diminish noise interference, and pay more attention to intrinsic RFF features, thereby facilitating effective extraction of RFF features and identification of wireless devices.

    \section{Proposed methodology}\label{sec:proposed-methodology}
    In this section, we present the details of the proposed method, comprising multiple periodic embedding representation, MPDFormer encoder, multi-periodicity feature adaptive-fusion, classifier, and loss functions.
    The overall architecture of the proposed method is depicted in Fig.~\ref{fig:total_network_architecture}.

    \subsection{Multi-periodicity embedding representation}

    \subsubsection{Spectrum offset}\label{subsec:spectrum-offset-&-multi-periodicity-processing}

    \begin{figure*} [h]
        \begin{center}
            \includegraphics[width=0.9\textwidth]{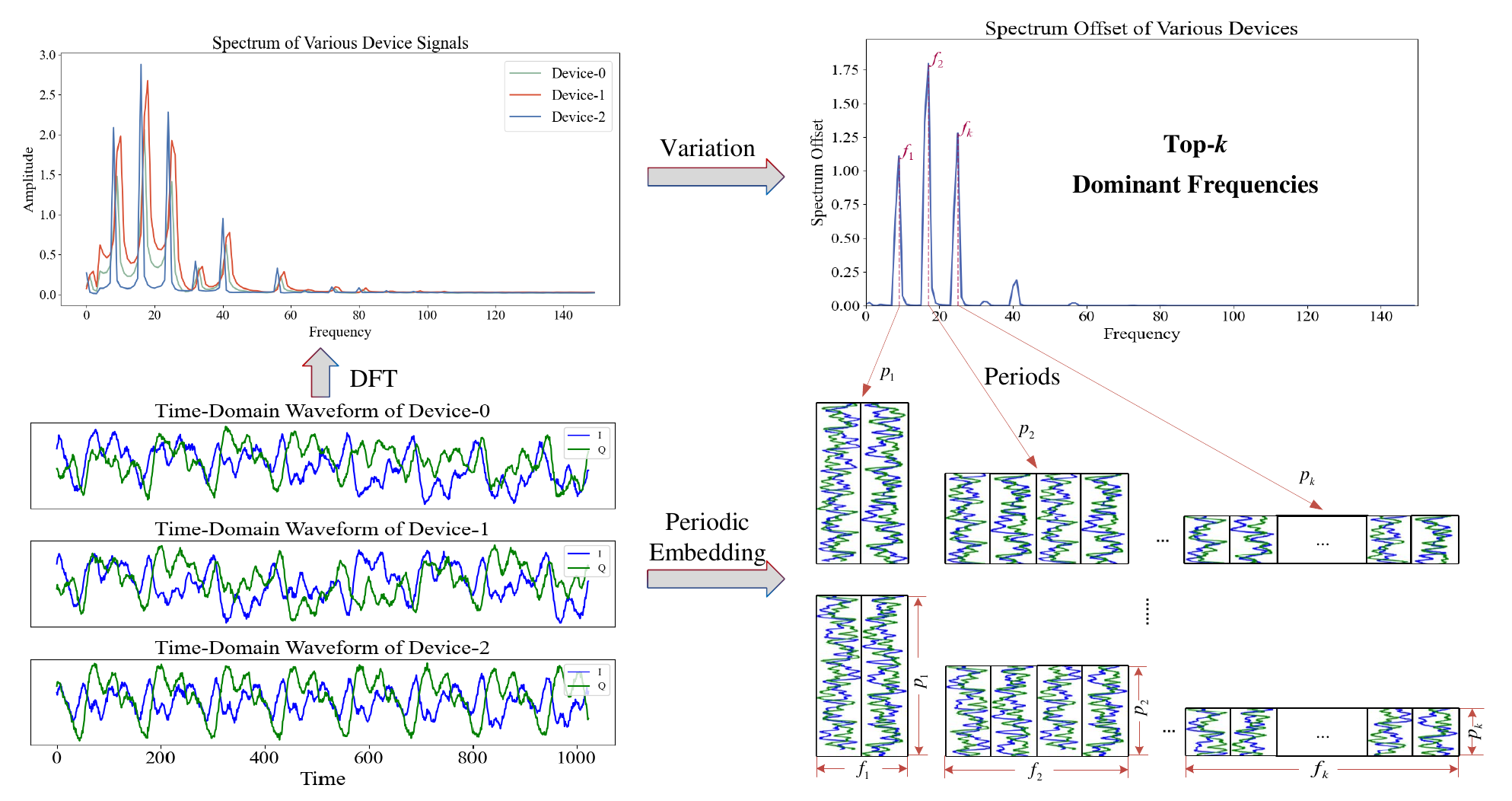}
        \end{center}
        \caption{Spectrum offset analysis and periodic embedding representation of RF signals.}
        \label{fig:multiperiod_from_spectrum_offset}
    \end{figure*}

    According to the previous study~\cite{wu2022DSLNSecuringInternet}, it is observed that the RFF features of the device encompass the internal thermal noise caused by electronic components, the nonlinearity distortion due to filters, frequency offsets because of oscillators, etc.
    These RFF features primarily manifest in the modulation property of RF signals, including amplitude offset, frequency offset, phase offset, etc.

    Given the modulated signal $x(t)$ and the transmitted signal $s(t)$ of a wireless device, the complex baseband signal of the received signal $r(t)$ is written as:
    \begin{equation}
        \label{eq:complex_baseband_of_received_signal}
        \begin{aligned}
            r(t)&=s(t)*\mathcal{C}(t)+\omega(t)\\
            &=x(t)*\mathcal{H}(t)*\mathcal{C}(t)+\omega(t),
        \end{aligned}
    \end{equation}
    where $*$ is the the convolution operation in the field of communication, $\mathcal{H}(t)$ denotes the impulse response of hardware impairments, $\mathcal{C}(t)$ represents the channel impulse response.
    In an ideal channel environment, devoid of hardware impairments, the spectrum characteristics of the received signal should be theoretically identical.
    In real-world wireless channel conditions that are uniform, we assume that all devices encounter analogous channel effects.
    Consequently, Eq~\eqref{eq:complex_baseband_of_received_signal} can be simplified to the impact of hardware impairments, which can be expressed as follows:
    \begin{equation}
        \label{eq:simple_complex_baseband_of_received_signal}
        \begin{aligned}
            r(t)=x(t)*\mathcal{H}(t)+\omega(t).
        \end{aligned}
    \end{equation}

    By performing a discrete Fourier transform on Eq.~\eqref{eq:simple_complex_baseband_of_received_signal}, the Power Spectral Density (PSD) of the received signal $r(t)$ can be expressed as:
    \begin{equation}
        \label{eq:fft_of_complex_baseband_of_received_signal}
        \begin{aligned}
            R(f)&= \mathcal{F}\{r(t)\}\\
            &= X(f) \cdot \mathbb{H}(f) + W(f),
        \end{aligned}
    \end{equation}
    where $\mathcal{F}(\cdot)$ denotes the DFT, $X(f)$, $\mathbb{H}(f)$ and $W(f)$ represent the PSD of the normalized modulated signal, hardware impairments, and additive white Gaussian noise, respectively.

    All ZigBee devices conform to the same protocol, thereby theoretically sharing the same signal attributes such as modulation type, bandwidth, carrier frequency, etc.
    Consequently, The PSD of different ZigBee signals should should approximately present similar results.
    However, due to the variability in RFF features, subtle differences are observed in the PSDs of various ZigBee signals.
    We refer to this type of variation as \textbf{spectrum offset}.

    \subsubsection{Multi-periodicity analysis}\label{subsec:multi-periodicity-processing}

    Based on the aforementioned analysis, we find the PSD differences under the impact of the spectrum offset, to derive the difference of RFF features in frequency domain.
    To find the peaks of spectrum offsets, we calculates the variation to describe the degree of deviation of the spectrum offset among all device signals.
    This serves as a measurement of the difference in PSD among these frequencies, thereby magnifying the spectrum offset caused by RFF features of various wireless devices.
    The spectrum offset is written as:
    \begin{equation}
        \label{eq:spectrum_offset_with_coefficient_variation}
        \begin{aligned}
            \bm{\mathbf{V}}(f) = \sqrt {E\{\mathbf{R}(f)-\boldsymbol{\mu}(f)\}},
        \end{aligned}
    \end{equation}
    where $\boldsymbol{\mu}(f)$ denotes the mean of the PSD among all devices, and can be expressed as:
    \begin{equation}
        \label{eq:mean_and_standard}
        \begin{aligned}
            \boldsymbol{\mu}(f) &= E\{\mathbf{R}(f)\}= \frac{1}{L}\sum_{n=0}^{L}R_{n}(f)+W_{n}(f),
        \end{aligned}
    \end{equation}
    where $L$ is the length of the input signal.
    Concerning the given discrete time-series signal ,the variance analysis based on the spectrum offset $\mathbf{V}(f)$ provides the following insights:
    \begin{equation}
        \label{eq:topk_frequency_period}
        \begin{aligned}
            \mathbf{\hat{A}}, \mathbf{\hat{F}}, \mathbf{\hat{P}} = \underset{f\subset\{1, 2, \cdots, L\}}{\mathrm{argtopk}}\{\mathbf{V}(f)\},
        \end{aligned}
    \end{equation}
    where
    \begin{equation}
        \label{eq:topk_three_elements}
        \begin{aligned}
            \mathbf{\hat{A}}&=\{\hat{a}_1, \hat{a}_2, \cdots, \hat{a}_k\},\\
            \mathbf{\hat{F}}&=\{\hat{f}_1, \hat{f}_2, \cdots, \hat{f}_k\},\\
            \mathbf{\hat{P}}&=\{\hat{p}_1, \hat{p}_2, \cdots, \hat{p}_k\},
        \end{aligned}
    \end{equation}
    where $a_k$ denotes the $k-th$ amplitude of the spectrum, $\hat{p}_k$ denotes the period corresponding to the $k-th$ frequency $\hat{f}_k$.
    The relationship between $\hat{f}_k$ and $\hat{p}_k$ can be represented as:
    \begin{equation}
        \label{eq:frquency_vs_period}
        \begin{aligned}
            \hat{p}_i &= \left \lceil \frac{L}{\hat{f}_i} \right \rceil, i\in \{1, 2, \cdots, k\},
        \end{aligned}
    \end{equation}
    where $\left \lceil \cdot \right \rceil$ represents rounding the result of the input upwards to the nearest integer.

    The top-$k$ frequencies corresponding to the dominant peak of spectrum offset yield the several frequencies exhibiting the maximum differences of RFF features and their corresponding periodic components.
    $k$ is a hyperparameter.
    Considering the spectral sparsity of time-series signals~\cite{subbareddy2021GraphLearningSpectral} and the characteristics of multi-periodicity in time-series signals~\cite{wu2023TimesNetTemporal2DVariation, wu2022AutoformerDecompositionTransformers}, we choose the top-$k$ dominant peaks of spectrum offset.
    To avoid the impact of meaningless noise at high-frequency components, the selection of $k$ is typically limited.

    \subsubsection{Periodic embedding representation}

    Building upon the findings of the top-$k$ spectrum offset peaks from the time-series signal analysis, the periods corresponding to the tok-$k$ dominant frequencies are calculated via Eq.~\eqref{eq:frquency_vs_period}.
    These periodic components present the temporal patterns inherent in time-series signals and encapsulate the maximum spectrum offset, epitomizing the primary differential components within the RFF features.
    Consequently, we attain a periodic embedding representation for RF signals by periodically intercepting these time-series signals.

    The representation achieved through periodic embedding serves a dual purpose.
    On one hand, it ensures the retention of the most integral RFF features of the time-series signal, surpassing the efficacy of frequency-based features from the respective of the frequency domain.
    On the other hand, the periodic embedding representation derived using spectrum offsets inherently encompasses the maximal periodic differentiation within the RFF features in the time domain.

    In accordance with Eq.~\eqref{eq:topk_frequency_period}, the length of all periodic embedding representations should align with the multiples of the head count within the Transformer-based multi-head attention mechanism, which typically adheres to a power of two.
    This alignment ensures divisibility by the head count.
    From an alternative perspective, fine-tuning for frequency and period is deemed necessary to counteract the deviations introduced by noise and channel effects.
    Fig.~\ref{fig:multiperiod_from_spectrum_offset} illustrates the specific calibration required for the periods and frequencies, which can be mathematically expressed as follows:
    \begin{equation}
        \label{eq:approximated_periods}
        \begin{aligned}
            \mathbf{F} &= f_1, f_2, \cdots, f_k = Refine(\hat{f}_1, \hat{f}_2, \cdots, \hat{f}_k),\\
            \mathbf{P} &= p_1, p_2, \cdots, p_k = Refine(\hat{p}_1, \hat{p}_2, \cdots, \hat{p}_k).
        \end{aligned}
    \end{equation}

    Consequently, for a one-dimensional (1D) time-series signal $\mathbf{R}_{1D} \subset \mathbb{R}^{L\times 2}$, where $L$ represents the length of the input signals, and $2$ denotes the In-phase (I) and Quadrature (Q) of the input signals,
    the two-dimensional (2D) output from the periodic embedding representation is derived as:
    \begin{equation}
        \label{eq:periodic_embedding}
        \begin{aligned}
            \mathbf{R}_{2D}^{i} = PER_{p_i, f_i}(\mathbf{R}_{1D}), i\in\{1, 2, \cdots, k\},
        \end{aligned}
    \end{equation}
    where $PER(\cdot)$ denotes the function for periodic embedding representation.

    Additionally, to preserve signal periodicity and ensure a balanced representation of the I and Q channels, the Q-channel signal is sequenced subsequent to the I-channel within the embeddings.
    This sequencing facilitates independent processing of periodic auto-correlation for each signal channel when divided by the head count.
    The 2D periodic embedding of the signal, inclusive of both IQ channels, is written as:
    \begin{equation}
        \label{eq:2d_periodic_embedding}
        \begin{aligned}
            \mathbf{R}_{2D}^{i} &= \{r_{I, 1}^i, r_{I, 2}^i, \cdots, r_{I, N_i}^i, r_{Q, 1}^{i},r_{Q, 2}^{i}, \cdots, r_{Q, N_i}^{i}\},\\
            \mathbf{R}_{2D} &= \{ \mathbf{R}_{2D}^{1}, \mathbf{R}_{2D}^{2}, \cdots, \mathbf{R}_{2D}^{k}\},
        \end{aligned}
    \end{equation}
    where $r_{I,n}^i$ and $r_{Q,n}^i$ signify the $i$-th period of the $n$-th I-channel and Q-channel signals respectively, $\mathbf{R}_{2D}^i \subset \mathbb{R}^{1 \times 2p_i}$ represents the $i$-th periodic signal embedding, and $\mathbf{R}_{2D} \subset \mathbb{R}^{k \times 2p_i}$ denotes the comprehensive periodic embedding representation of the signal, with $N_i=\frac{L}{p_i}$ indicating the count of periodic embeddings in the $i$-th period $p_i$.

    \subsection{MPDFormer Encoder}

\begin{figure*} [h]
        \begin{center}
            \includegraphics[width=1.0\textwidth]{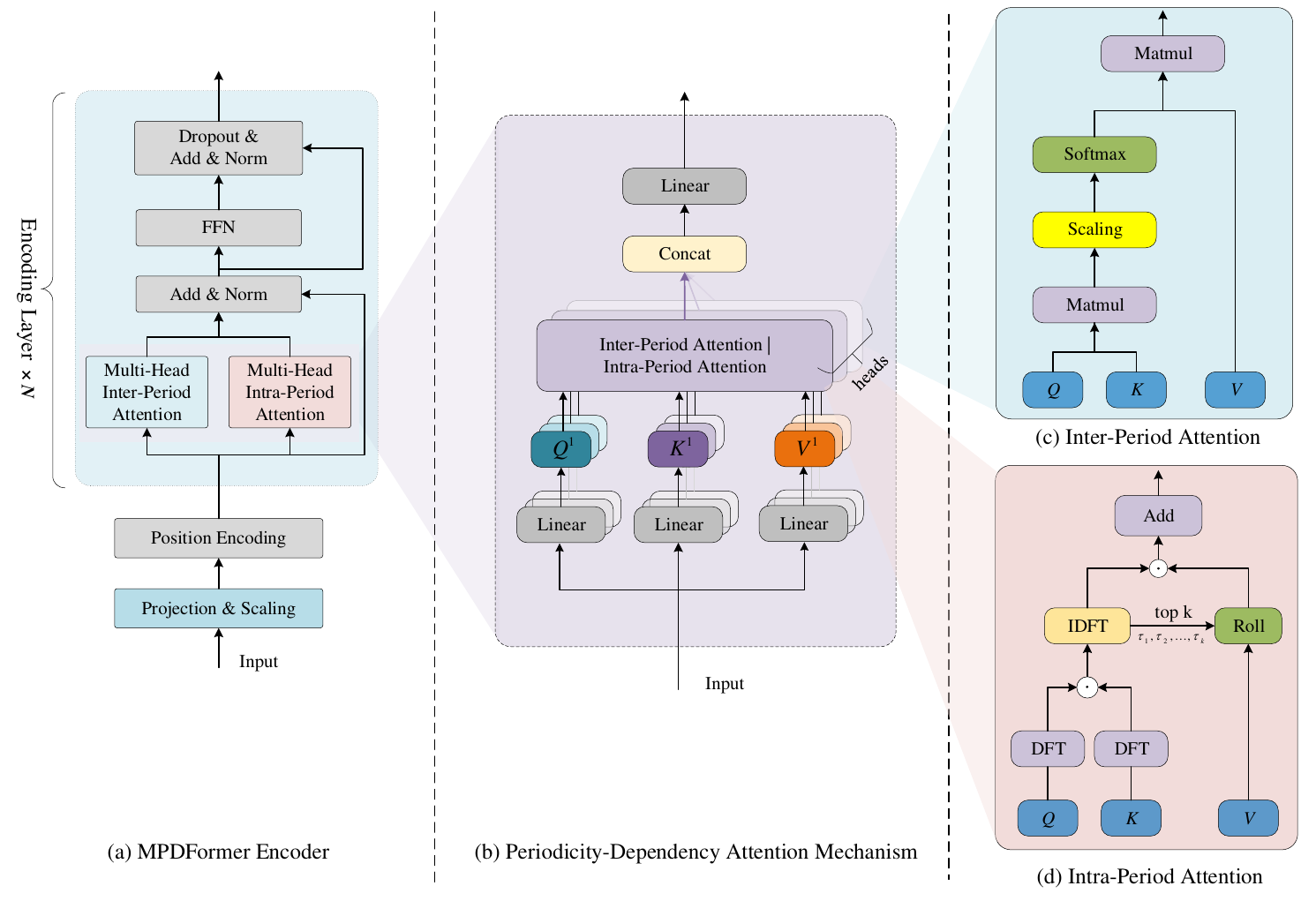}
        \end{center}
        \caption{Network architecture of the MPDFormer encoder and its constituent modules. (a) Architecture of the MPDFormer encoder. (b) Periodicity dependency attention mechanism. (c) Inter-period attention. (d) Intra-period attention. Projection refers to a linear operation applied to input signals for linear space transformation, $|$ signifies the 'or' operation, and $\odot$ denotes the Hadamard Product.}
        \label{fig:MPDFormer_Encoder}
    \end{figure*}

    This subsection presents the architecture of the multi-periodicity dependency Transformer encoder, which primarily comprises positional encoding, periodicity-dependency attention mechanism, feed forward network.
    the network architecture of the MPDFormer encoder is shown in Fig.~\ref{fig:MPDFormer_Encoder}

    \subsubsection{Positional encoding}

    In the MPDFormer architectures, positional encoding is an indispensable component for the interpretation of sequential signals.
    Transformers, in contrast to recurrent neural network architectures such as RNNs or LSTMs, lack an intrinsic mechanism to process sequential signal, owing to their reliance on the self-attention mechanism.
    This mechanism empowers MPDFormer to selectively focus on various frames within the input signal for each output computation, thereby relinquishing any inherent positional or sequential context.
    To counterbalance this limitation, positional encoding is employed to infuse information regarding the position of frames within the input signal.
    The formulation of positional encoding is expressed as:
    \begin{equation}
        \label{eq:the_original_position_encoding_of_Transformer}
        \begin{aligned}
            &\mathbf{PE}_{p_i, 2i} = \sin(pos/10000^{i/{p_i}}),\\
            &\mathbf{PE}_{p_i, 2i+1} = \cos(pos/10000^{i/{p_i}}),
        \end{aligned}
    \end{equation}
    where $pos$ symbolizes the position of the periodic embeddings of the signal, and $2i$ and $2i+1$ represent the ($2i$)-th and ($2i+1$)-th positions, respectively.

    Subsequently, these positional encodings are aggregated with the inputs $\mathbf{R}_{2D}^{i}$, culminating in the output of positional encodings as written by:
    \begin{equation}
        \label{eq:positional_encodings}
        \begin{aligned}
            \widehat{\mathbf{R}}_{2D}^{i} = \mathbf{R}_{2D}^{i} + \mathbf{PE}_{p_i},
        \end{aligned}
    \end{equation}
    where $\mathbf{PE}_{p_i}$ is computed in accordance with Eq.~\eqref{eq:the_original_position_encoding_of_Transformer}.

    \subsubsection{Periodicity-dependency attention mechanism}

    The MDPFormer framework depicted in Fig.~\ref{fig:MPDFormer_Encoder} introduces the periodicity-dependency attention mechanism which includes inter-period and intra-period attention mechanisms, aiming to capture the long-range RFF features from the correlation within and across the periodic embedding representation.
    Initially, the signal undergoes a projection and scaling transformation, refining its feature space to enhance data distribution characteristics.
    Sequentially, the model incorporates position encoding, intrinsic to MPDFormer architectures, imparting sequential order information and enabling the discernment of positions within the periodic representations of the input signal.

    Subsequently, the MPDFormer's encoding layers, iteratively applied $N$ times subsequent to the position encoding operation, function to transmute the input periodic embedding into a contextually enriched representation, thereby augmenting the proposed model for downstream tasks.
    As indicated in Fig.~\ref{fig:MPDFormer_Encoder}a, we supersede the self-attention mechanism and residual connections of the conventional MPDFormer with our proposed inter-period and intra-period attention mechanisms.
    The inter-period attention is designed to perform long-range attention calculations across periodic embeddings, capturing dependencies over periods by the inter-period attention.
    Conversely, intra-period attention introduces a varied temporal delay to the periodic embeddings, thereby enabling auto-correlation calculations within a single period to unravel temporal correlations therein.

    The intricate architecture of the inter-period and intra-period attention mechanisms is presented in Fig.~\ref{fig:MPDFormer_Encoder}b.
    Considering an input signal $\mathbf{Y}_{1D}^i(t)$ transformed into a periodic embedding representation $\mathbf{Y}_{2D}^i(t)$, the associated auto-correlation matrix is expressed as:
    \begin{equation}
        \label{eq:time_delay_correlation}
        \begin{aligned}
            \mathbf{C}_{YY}(\tau)&= \lim_{L \to \infty} \frac{1}{L} \sum_{t=1}^{L} \mathbf{Y}_{2D}^i(t) \mathbf{Y}_{2D}^i(t-\tau),
        \end{aligned}
    \end{equation}
    where $\tau$ signifies the temporal delay.
    Given the periodic embedding representations $\mathbf{Q}_{2D}^i$ and $\mathbf{K}_{2D}^i$  with specified period $p_i$, we, motivated by previous work~\cite{vaswani2017AttentionAllYou}, ascertain the period-delay correlation as:
    \begin{equation}
        \label{eq:period_delay_correlation}
        \begin{aligned}
            \mathbf{C}_{QK, Inter}(p_i)&=E\{\mathbf{Q}_{2D}^i(t) [\mathbf{K}_{2D}^i(t-p_i)]^T\}\\
            &=\begin{bmatrix}
                  \mathbf{C}_{Q_1, K_1}     & \cdots & \mathbf{C}_{Q_1, K_{N_i}}     \\
                  \vdots                    & \ddots & \vdots                        \\
                  \mathbf{C}_{Q_{N_i}, K_1} & \cdots & \mathbf{C}_{Q_{N_i}, K_{N_i}}
            \end{bmatrix},
        \end{aligned}
    \end{equation}
    where $T$ is the transpose operation. The element in Eq.~\eqref{eq:period_delay_correlation} can be expressed as:
    \begin{equation}
        \label{eq:period_delay_correlation_elements}
        \begin{aligned}
            \mathbf{C}_{Q_m, K_n} = \frac{1}{p_i} \sum_{i=0}^{p_i} q_{i-m \cdot p_i}  \cdot  k_{i-n \cdot p_i}, m, n \in \{0, 1, \cdots, N_i\},
        \end{aligned}
    \end{equation}
    where $N_i = \frac{L}{p_i}$, and $L$ representing the length of the given input signal $\mathbf{R}_{1D} \subset \mathbb{R}^{L\times 2}$.
    Owing to the signal embedding with a period $p_i$, the inter-period attention mechanism is characteristically delayed by the period.
    The mechanism is mathematically formulated as:
    \begin{equation}
        \label{eq:inter_period_attention_mechanism}
        \begin{aligned}
            \Lambda^{i} &= IEP(\mathbf{Q}_{2D}^i, \mathbf{K}_{2D}^i, \mathbf{V}_{2D}^i) \\
            &= softmax({\mathbf{C}_{QK, Inter}(p_i)})\mathbf{V}_{2D}^i,
        \end{aligned}
    \end{equation}
    where $IEP(\cdot)$ denotes the IntEr-Period attention function.
    This attention mechanism lays the groundwork for unearthing long-range dependencies among periodic or repetitive RFF features across wireless devices.

    To further capture the intra-period RFF features, we present an intra-period attention mechanism.
    In accordance with the Wiener-Khinchin theorem~\cite{wiener1964TimeSeries}, the auto-correlation of $\mathbf{Y}_{2D}^{i}(t)$ can be determined by utilizing its PSD.
    This theorem can be extended to calculation of the cross correlation with $\mathbf{Q}_{2D}^{i}(t)$ and $\mathbf{K}_{2D}^{i}(t)$ derived from the same input $\mathbf{R}_{2D}^{i}$ by the motivation of previous work~\cite{wu2022AutoformerDecompositionTransformers}, which is expressed as:

    \begin{equation}
        \label{eq:correlation_of_Intra}
        \begin{aligned}
            \mathbf{C}_{QK, Intra}(\tau) &= \mathcal{F}^{-1} ( \mathbf{S}_{QK, Intra}(f))\\
            &=\mathcal{F}^{-1}(\mathcal{F}(\mathbf{Q}_{2D}^{i})\mathcal{F}^{*}(\mathbf{K}_{2D}^{i})),
        \end{aligned}
    \end{equation}
    where $\mathcal{F}^{-1}(\cdot)$ denotes the inverse DFT, $\mathcal{F}^{*}(\cdot)$ is the conjugate operation of DFT, $\mathbf{S}_{QK, Intra}(f)$ is the cross PSD of the cross-correlation of $\mathbf{Q}_{2D}^{i}(t)$ and $\mathbf{K}_{2D}^{i}(t)$.
    Commencing with the periodic embedding representation, we compute the maximum correlation with the most substantial temporal delay for $\tau < p_i$ as follows:
    \begin{equation}
        \label{eq:maximum_correlation}
        \begin{aligned}
            \tau_1, \tau_2, \cdots, \tau_k = \underset{\tau\subset\{1, 2, \cdots, p_i\}}{\mathrm{argtopk}}\{\mathbf{C}_{QK, Intra}(\tau)\}.
        \end{aligned}
    \end{equation}

    Thus, the intra-period attention with the temporal delay characteristic is expressed as:
    \begin{equation}
        \label{eq:temporal_period_delay_correlation}
        \begin{aligned}
            \Sigma^{i} &= IAP(\mathbf{Q}_{2D}^i, \mathbf{K}_{2D}^i, \mathbf{V}_{2D}^i) \\
            &=\sum_{i=1}^{k} \mathcal{R}oll(\mathbf{V}_{2D}^i) \cdot softmax(\mathbf{C}_{QK, Intra}(\tau_i)),
        \end{aligned}
    \end{equation}
    where $IAP(\cdot)$ represents the IntrA-Period attention function, and $\mathcal{R}oll(\cdot)$ effectuates a cyclic shift of the vector elements.
    This operation culminates in the accentuation of RFF features correlated within a single period, significantly improving the discernment of RFF features.

    To fuse the output of these two attention mechanisms, inter-period and intra-period attentions are weighted by the balance factors.
    The fusion operation can be written as:
    \begin{equation}
        \label{eq:adaptively_fusing_attention}
        \begin{aligned}
            \varGamma=FA(\Lambda, \Sigma) = \mathbf{X} + \sigma \Lambda + \varsigma \Sigma,
        \end{aligned}
    \end{equation}
    where $FA(\cdot)$ denotes the fused attention, $\mathbf{X}$ is the input feature maps for residual connection, $\sigma$ and $\varsigma$ are the weight for balancing the inter-period and intra-period attention.

    \subsubsection{Feed-forward network}
    The Feed-Forward Network (FFN) emerges as a pivotal component within the encoder architecture, strategically positioned subsequent to the periodicity-dependency attention mechanism.
    It fulfills a crucial role in computing the final representation of each signal.
    Distinct from the periodicity-dependency attention mechanism, the FFN operates on an individual basis for each positional input, thereby treating the input at each position in an isolated manner.
    Functioning as a position-wise, fully-connected layer, the FFN uniformly applies a predefined learned linear transformation across each positional input.
    It serves to augment the periodicity-dependency attention mechanism, introducing an additional layer of abstraction that aids in deciphering intricate patterns and correlations inherent within each signal's periodic embedding representation.

    From a mathematical perspective, the FFN is composed of two linear transformations interspersed with a Rectified Linear Unit (ReLU) activation.
    The two linear transformations are parameterized by weight matrices $\mathbf{W}_1 \in \mathbb{R}^{ 2p_i \times N_i}$ and $\mathbf{W}_2 \in \mathbb{R}^{ N_i \times 2p_i}$, and bias terms $\mathbf{b}_1 \in \mathbb{R}^{2p_i}$ and $\mathbf{b}_2 \in \mathbb{R}^{N_i}$.
    For a given input $\varGamma^{i} \in \mathbb{R}^{N_i \times 2p_i}$, the FFN is computed as:
    \begin{equation}
        \label{eq:ffn}
        \begin{aligned}
            FFN(\varGamma^{i}) = \rho(0, \mathbf{\varGamma}^{i} \mathbf{W}_1 + \mathbf{b}_1)\mathbf{W}_2 + \mathbf{b}_2,
        \end{aligned}
    \end{equation}
    where $\rho$ denotes the ReLU activation function.
    Eq.~\eqref{eq:ffn} encapsulates the FFN's operational sequence: an initial linear transformation of the input, succeeded by a ReLU activation, and culminating in a subsequent linear transformation.
    The output dimensionality of the FFN mirrors that of its input, ensuring seamless integration into the MPDFormer's subsequent layers.

    \subsection{Multi-periodicity feature adaptive-fusion}
    In the pursuit of deducing the output probability distribution, the encoders of MPDFormer engage the input $\mathbf{R}_{2D}^i \subset \mathbb{R}^{1 \times 2p_i}$.
    The overall procedural calculation is expressed as follows:
    \begin{equation}
        \label{eq:output_from_MPDFormer_encoder}
        \begin{aligned}
            \mathbf{E}^1, \mathbf{E}^2, \cdots, \mathbf{E}^i &= Encoder(\mathbf{R}_{2D}^1, \mathbf{R}_{2D}^2, \cdots, \mathbf{R}_{2D}^i),\\
            \mathbf{T}^1, \mathbf{T}^2, \cdots, \mathbf{T}^i &= Transpose(\mathbf{E}^1, \mathbf{E}^2, \cdots, \mathbf{E}^i),\\
            \mathbf{G}^1, \mathbf{G}^2, \cdots, \mathbf{G}^i &= GAP(\mathbf{T}^1, \mathbf{T}^2, \cdots, \mathbf{T}^i),\\
        \end{aligned}
    \end{equation}
    wherein $GAP(\cdot)$ signifies global average pooling, and $N_{max}$ denotes the maximal entity among all $N_i$ corresponding to the period $p_i$.

    To efficaciously harness multi-periodicity RFF features for the subsequent classification task of various devices, a fusion of the output feature maps from various periodicity-dependency encoders of MPDFormer is imperative.
    Building upon insights gained from our previous study~\cite{xiao2023MultiscaleCorrelationNetworks}, we introduce a adaptive weighting scheme tailored for arbitrary feature dimensions incorporating with a padding operation, which is expressed as:
    \begin{equation}
        \label{eq:padding_operation}
        \begin{aligned}
            \hat{\mathbf{G}}^1, \hat{\mathbf{G}}^2, \cdots, \hat{\mathbf{G}}^i = Padding_{N_{max}}(\mathbf{G}^1, \mathbf{G}^2, \cdots, \mathbf{G}^i).\\
        \end{aligned}
    \end{equation}
    This scheme is specifically designed to equilibrate the multiple periodicity-dependency RFF Features, thereby enhancing the representational efficacy of these features.
    The adaptive weighting procedure is written as:
    \begin{equation}
        \label{eq:adaptive_weighting}
        \begin{aligned}
            \mathbf{W}^i &= WA(\hat{\mathbf{G}}^i) \\
            &= \varrho (\mathbf{W}_{2} \delta (\frac{\mathbf{W}_{1}}{T} \sum_{m=1}^{N_{max}}\sum_{n=1}^{2p_i} \hat{\mathbf{G}}^i + \mathbf{b}_{1}) + \mathbf{b}_{2}),
        \end{aligned}
    \end{equation}
    where $\varrho$ is a sigmoid function, $\delta$ represents the SeLU function, $p_i$ represent the $i$-th period of the periodic embedding representation, $\mathbf{W}_{1}$ and $\mathbf{W}_{2}$ are learnable weights.
    Thus, the adaptive fusion of multi-periodicity RFF features is expressed as:
    \begin{equation}
        \label{eq:multi_period_feature_fusion}
        \begin{aligned}
            \mathcal{\mathbf{W}}^1, \mathcal{\mathbf{W}}^2, \cdots, \mathcal{\mathbf{W}}^i &= WA(\hat{\mathbf{G}}^1, \hat{\mathbf{G}}^2, \cdots, \hat{\mathbf{G}}^i).
        \end{aligned}
    \end{equation}

    Subsequently, these weighted feature maps undergo concatenation to consummate the weighted fusion of multi-periodicity RFF features.
    The output is written as:
    \begin{equation}
        \label{eq:feature_fusion}
        \begin{aligned}
            \mathcal{\mathbf{O}} &= Concat(\mathcal{\mathbf{W}}^1, \mathcal{\mathbf{W}}^2, \cdots, \mathcal{\mathbf{W}}^i).
        \end{aligned}
    \end{equation}

    \subsection{Classifier}

    In the illustrated architecture shown in Fig.~\ref{fig:total_network_architecture}, the classifier is a Multi-Layer Perceptron (MLP) designed to classify inputs into one of 32 potential categories.
    It includes three fully connected layers with decreasing sizes of 128, 64, and 32 units, respectively.
    Each layer successively refines features extracted from the input, culminating in a probability distribution over the label space, where each output node represents the likelihood of a specific class.

    Therefore, the network architecture of MPDFormer is shown in Table~\ref{tab:layout_of_network_architecture}.
    \begin{table}[htbp]
        \small
        \centering
        \caption{Layout of MPDFormer network architecture for RDR dataset.}
        \begin{tabularx}{\textwidth}{c|c|c|c}
            \hline
            \textbf{Layers}                   & \textbf{Output Dimension} & \textbf{Parameters} & \textbf{kFLOPs}          \\
            \hline
            Input                             & 2048 $\times$ 2           & --                  & --                       \\
            \hline
            Periodic Embedding Representation & 29 $\times$ 144           & --                  & \multirow{8}{*}{30515.5} \\
            \cline{1-3}
            Projection                        & 29 $\times$ 144           & 20,880              &                          \\
            \cline{1-3}
            Scaling                           & 29 $\times$ 144           & --                  &                          \\
            \cline{1-3}
            Positional Encoding               & 29 $\times$ 144           & --                  &                          \\
            \cline{1-3}
            Encoding Layers                   & 29 $\times$ 144           & 1,046,466           &                          \\
            \cline{1-3}
            Transpose                         & 144 $\times$ 29           & --                  &                          \\
            \cline{1-3}
            GAP                               & 144 $\times$ 1            & --                  &                          \\
            \cline{1-3}
            Squeeze                           & 144                       & --                  &                          \\
            \hline
            Periodic Embedding Representation & 37 $\times$ 112           & --                  & \multirow{8}{*}{24165.4} \\
            \cline{1-3}
            Projection                        & 37 $\times$ 112           & 12,656              &                          \\
            \cline{1-3}
            Scaling                           & 37 $\times$ 112           & --                  &                          \\
            \cline{1-3}
            Positional Encoding               & 37 $\times$ 112           & --                  &                          \\
            \cline{1-3}
            Encoding Layers                   & 37 $\times$ 112           & 635,454             &                          \\
            \cline{1-3}
            Transpose                         & 112 $\times$ 37           & --                  &                          \\
            \cline{1-3}
            GAP                               & 112 $\times$ 1            & --                  &                          \\
            \cline{1-3}
            Squeeze                           & 112                       & --                  &                          \\
            \hline
            Concatenation                     & 256                       & --                  & --                       \\
            \hline
            Adaptive Fusion                   & 256                       & 64                  & 0.06                    \\
            \hline
            Linear                            & 128                       & 32,896              & \multirow{3}{*}{32.8}    \\
            \cline{1-3}
            SeLU                              & 128                       & --                  &                          \\
            \cline{1-3}
            Dropout                           & 128                       & --                  &                          \\
            \hline
            Linear                            & 64                        & 8,256               & \multirow{3}{*}{8.2}     \\
            \cline{1-3}
            SeLU                              & 64                        & --                  &                          \\
            \cline{1-3}
            Dropout                           & 64                        & --                  &                          \\
            \hline
            Linear                            & 32                        & 2,080               & 2.0                      \\
            \hline
        \end{tabularx}

        \label{tab:layout_of_network_architecture}%
    \end{table}%

    \section{Signal acquisition and collection}\label{sec:signal-acquisition-and-collection}

    \begin{figure}[h]
        \centering
        \begin{minipage}{0.45\textwidth}
            \centering
            \includegraphics[width=\linewidth]{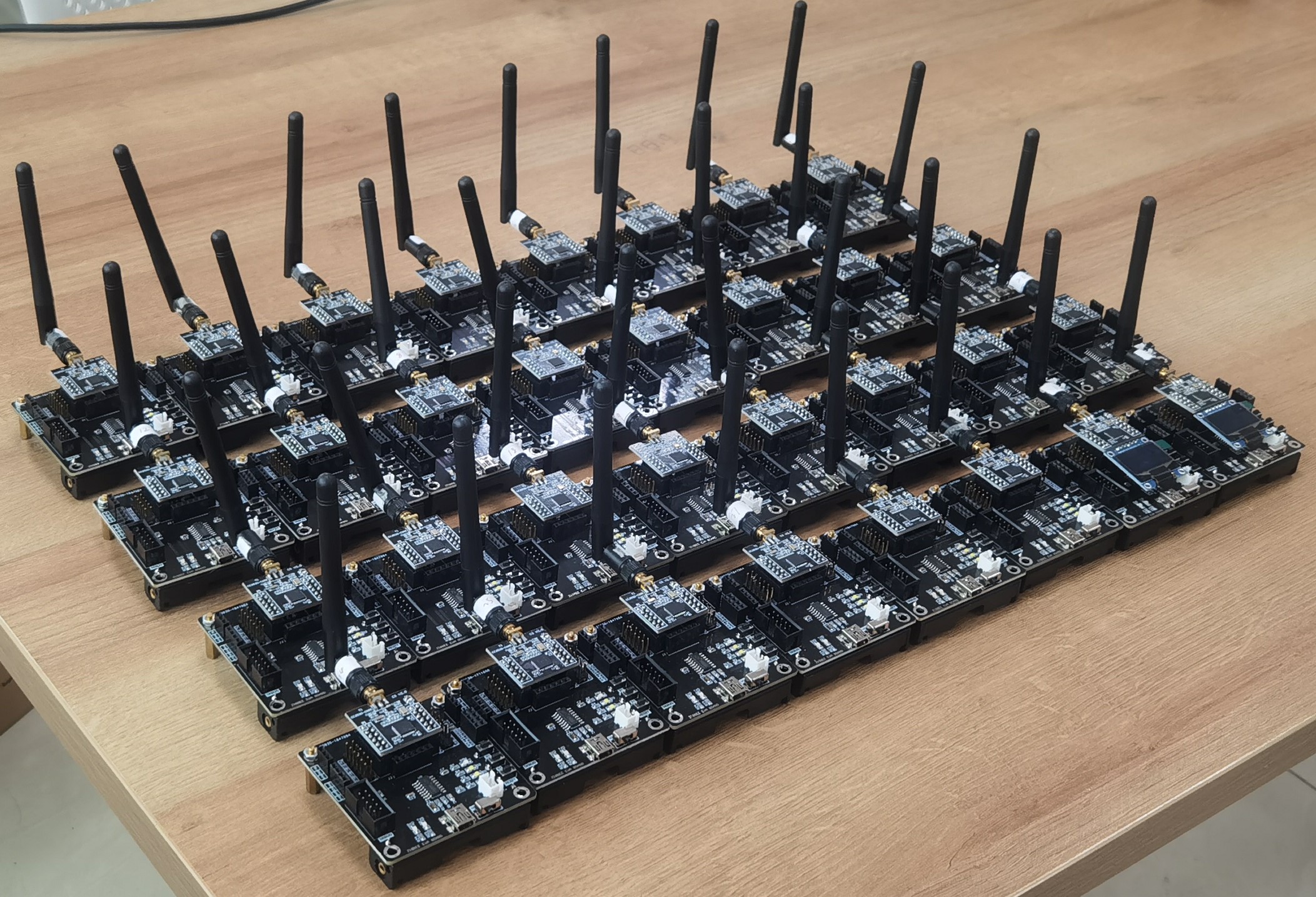}
            \subcaption{ZigBee Devices}
            \label{fig:ZigBee}
        \end{minipage}
        \hspace{0.1pt}
        \begin{minipage}{0.225\textwidth}
            \centering
            \includegraphics[width=\linewidth]{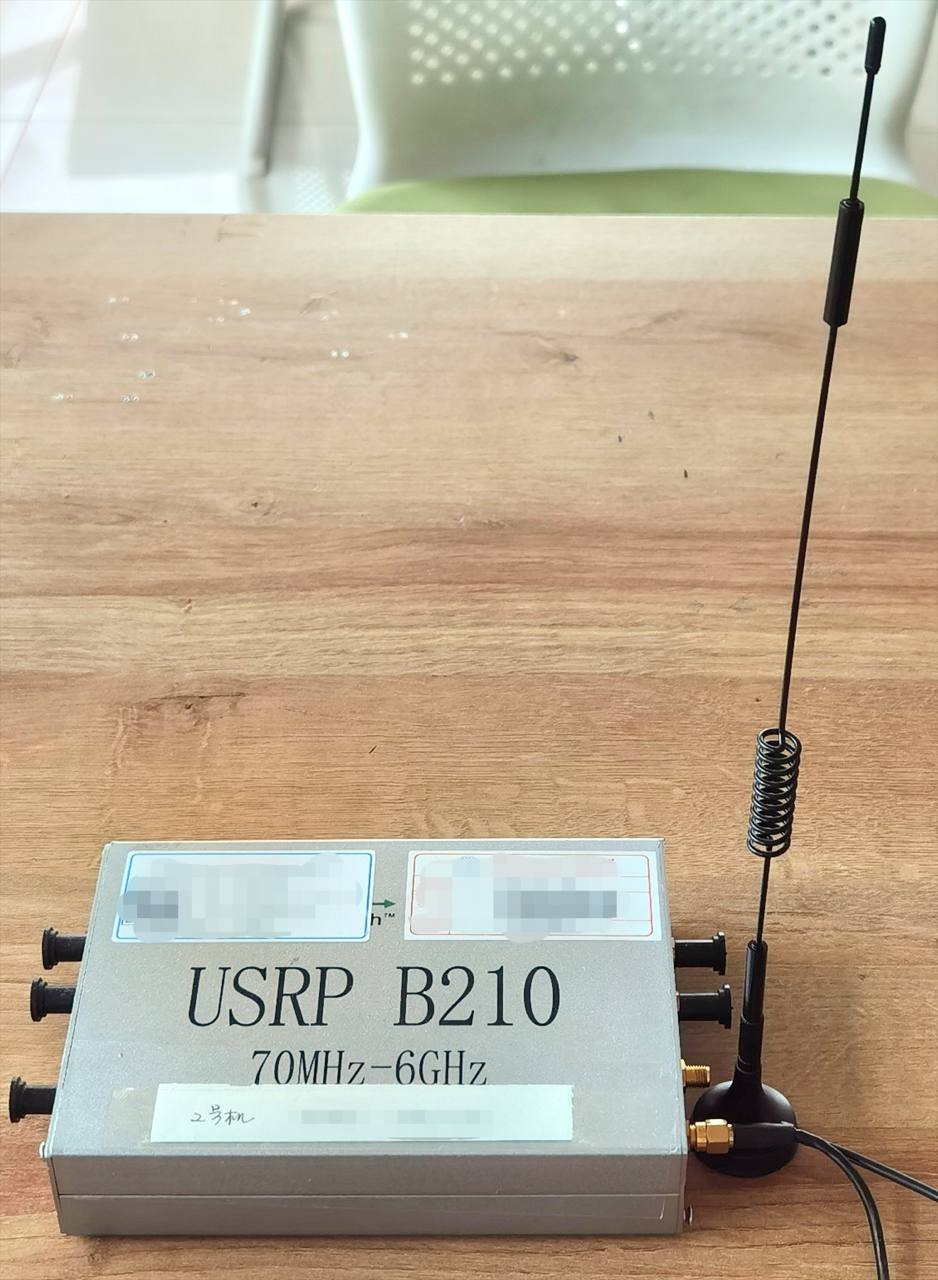}
            \subcaption{USRP B210}
            \label{fig:usrp_b210}
        \end{minipage}
        \caption{The signals are transmitted by 32 ZigBee devices and collected by an USRP B210 device.}
        \label{fig:hardwares}
    \end{figure}

    This section introduces two datasets of ZigBee signals pivotal for validating the proposed method.
    Subsequently, we present the characteristics of devices based on the ZigBee protocol, the hardware configurations of ZigBee devices, the software defined radio configurations of Universal Software Radio Peripheral (USRP) for capturing ZigBee signals, and the salient settings of the captured ZigBee signal datasets.

    Guided by the IEEE 802.15.4 standard, ZigBee represents a fundamental model for low-power, high-reliability wireless communications.
    Utilizing the Offset Quadrature Phase-Shift Keying (OQPSK) modulation, it incorporates the CC2530 module and operates within the 2.4GHz frequency band.
    This band is notable for its extensive industrial utility and lack of licensing requirements.
    With a transmission rate of 250kbps, ZigBee stands at the forefront of communication technologies, especially in sectors such as smart grid management, home automation, and healthcare monitoring.
    Its mesh networking features of ZigBee enable the integration of multiple devices, thus strengthening the infrastructure for future IoT advancements, underscoring ZigBee's significance in the technological realm.

    To simulate scenarios of unauthorized device infiltration in a ZigBee communication network, 32 devices were programmed to function as terminal nodes, as shown in Fig.~\ref{fig:ZigBee}.
    Unique ZigBee signal identifiers, such as destination and source addresses, were masked by developing firmware in C language using the IAR Embedded Workbench.
    Given that the collection of a ZigBee protocol terminal signal necessitates the presence of a coordinate node, we have executed hardware programming to circumvent the influence of the coordinate node on the terminal node collection.
    This enables the emission of ZigBee signals at a frequency of once per second without the requirement for self-organizing network.
    To minimize interference from other RF signals like Wi-Fi and Bluetooth, the carrier frequency of ZigBee signal was strategically set at 2.475GHz (channel 25).
    Moreover, the information and device identifiers were randomized within the Media Access Control (MAC) payload data bytes.

    \begin{figure*} [t]
        \begin{center}
            \includegraphics[width=1.0\textwidth]{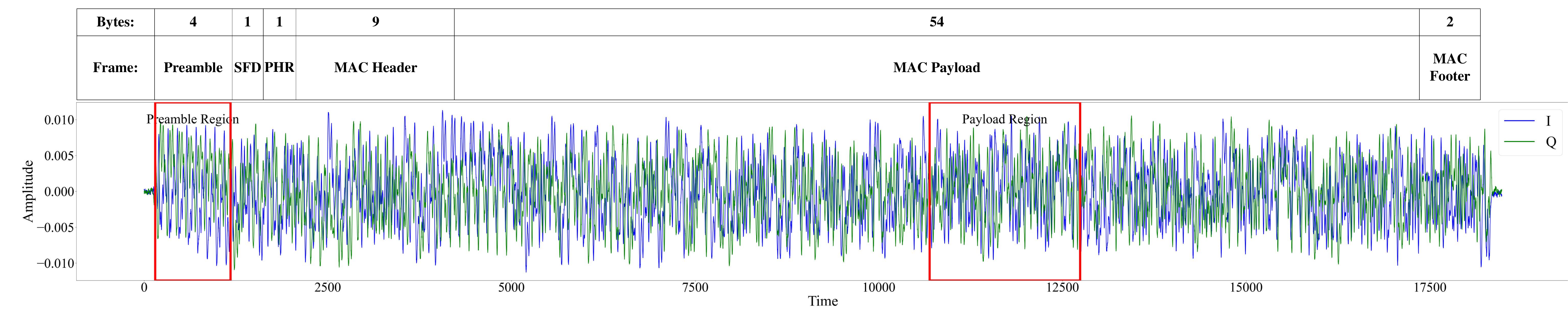}
        \end{center}
        \caption{Illustration of a ZigBee signal presents the preamble and payload regions. SFD denotes the start of frame delimiter, PHR denotes the physical layer header, while MAC denotes the media access control. The signal within the preamble region encompasses transient, semi-steady, and steady states, whereas the payload region is characterized by a steady signal.}
        \label{fig:ZigBee_signal}
    \end{figure*}

    For the purpose of signal acquisition, the research utilized an Ettus USRP B210 device equipped with a suction-cup antenna, as depicted in Fig.\ref{fig:usrp_b210}.
    This system, operating at a sampling rate of 8 MSamples/s, was selected to emulate the average system capabilities encountered in real-world scenarios.
    GNU Radio served as the software framework for signal processing, deliberately configured to operate without automatic gain control, IQ imbalance compensation, and direct current spike correction, to preserve the integrity characteristics of the raw signal.
    The structural composition of the data packet, as illustrated in Fig.\ref{fig:ZigBee_signal}, encompasses a 4-byte preamble, a 1-byte SFD, a 1-byte PHR, a 13-byte MAC header, a 54-byte MAC payload, and a 2-byte MAC footer.
    This configuration presents the intricate details of the ZigBee signal structure integral to the research.

    As depicted in Fig.~\ref{fig:ZigBee_signal}, this study presented two distinct signal datasets predicated upon their respective regions: the Constant Data Region (CDR) with same data bytes and Random Data Region (RDR) without same data bytes.
    The CDR signal encompasses transient, semi-steady, and steady-state segments within the preamble, whereas the RDR signal solely comprises a steady-state segment located within the MAC payload.

    \begin{table}[h]
        \small
        \centering
        \caption{Comparative Properties of CDR and RDR Datasets.}
        \begin{tabular}{c|p{2.5cm}p{2.5cm}}
            \toprule
            \multirow{2}[4]{*}{\textbf{Property}} & \multicolumn{2}{c}{\textbf{Dataset}} \\
            \cmidrule{2-3} & \centering \textbf{CDR} & \centering \textbf{RDR} \tabularnewline
            \midrule
            Center Frequency & \multicolumn{2}{c}{2.475GHz} \\
            Sampling Rate & \multicolumn{2}{c}{8 MSample/s} \\
            Categories & \multicolumn{2}{c}{32} \\
            Train:Test Ratio & \multicolumn{2}{c}{9:1} \\
            Data Byte & \centering Same & \centering Random \tabularnewline
            Dimension & \centering 1024$\times$2 & \centering 2048$\times$2 \tabularnewline
            SNRs & \multicolumn{2}{c}{from -20dB to 20dB with 2dB intervals} \\
            \bottomrule
        \end{tabular}%
        \label{tab:dataset_info}%
    \end{table}%

    In July 2023, a comprehensive RF signal collection was conducted for both datasets.
    Subsequently, during the signal pre-processing phase, the dimensionality of each signal was preserved at 1024$\times$2 and 2048$\times$2 for the CDR and RDR datasets, respectively.
    This process utilized 32 devices to represent the 32 types of classes within the ZigBee signal datasets.
    By maintaining the signal's length at 1024/2048 and designating for IQ channels, the signals' authenticity was safeguarded, yielding approximately 21,000 points per sample.
    The signals were preserved in their pristine form by eschewing synchronization measures and minimizing superfluous processing operations, thereby retaining native inconsistencies such as frequency offsets, phase offsets, and time shifts of which characteristics intrinsic to the wireless channel and RFF features.
    To emulate diverse noise environments, additive white Gaussian noise (AWGN) was deliberately introduced into varying samples, encompassing SNRs ranging from -20dB to 20dB at a 2dB interval.
    The datasets are split into training and testing sets at a ratio of 9:1.
    The property settings of the two datasets are presented in Table.~\ref{tab:dataset_info}.

    \section{Experiment and results}\label{sec:experiment-and-results}

    \subsection{Baseline methods}

    In the domain of RFFI, DL-based RFFI methods are primarily categorized into three DL framework: Convolutional Neural Networks (CNNs), Recurrent Neural Networks (RNNs), and Transformer-based frameworks~\cite{shen2023LengthVersatileNoiseRobustRadio}.
    Within the domain of CNNs, the ResNet architecture has been established as an exemplary model, primarily for its adeptness in hierarchical feature extraction.
    This deep neural network architecture is particularly renowned for its effective mitigation of gradient vanishing and explosion, which are common pitfalls in deep learning models.
    Pertaining to the processing of sequential data, RNNs and their derivatives, Long Short-Term Memory networks (LSTMs) and Gated Recurrent Unit networks (GRUs), have demonstrated their prowess.
    These architectures adeptly navigate the temporal intricacies inherent in RF signals, addressing salient the challenge of the management of long-range dependencies.
    The Transformer architecture, renowned for its self-attention mechanism~\cite{vaswani2017AttentionAllYou}, introduces a paradigm shift in recognizing and capitalizing on long-range correlations within RFF features, thereby achieving an unparalleled level of pattern recognition fidelity.
    Collectively, these DL-based methods have substantially expanded the research horizons in the RFFI domain, engendering a new epoch of precision and innovation in wireless device identification.

    \begin{table}[htbp]
        \centering
        \small
        \caption{Hyperparameters of various baseline methods.}
        \resizebox{\columnwidth}{!}{
            \begin{tabular}{c|ccccc}
                \toprule
                \textbf{Hyperparameters} & \textbf{GLFormer}~\cite{deng2023LightweightTransformerBasedApproach} & \textbf{Transformer}~\cite{vaswani2017AttentionAllYou} & \textbf{GRU}~\cite{shen2023LengthVersatileNoiseRobustRadio} & \textbf{LSTM}~\cite{shen2023LengthVersatileNoiseRobustRadio} & \textbf{ResNet18}~\cite{he2016DeepResidualLearning} \\
                \midrule
                Number of Layers         & 5                                                                    & 5                                                      & 2                                                           & 2                                                            & 4                                                   \\
                Heads                    & /                                                                    & 4                                                      & /                                                           & /                                                            & /                                                   \\
                Dimension                & 64                                                                   & 64                                                     & 64                                                          & 64                                                           & 1024/2048                                           \\
                Hidden Size/Channel      & /                                                                    & /                                                      & 256                                                         & 256                                                          & \{64, 128, 256, 512\}                               \\
                \bottomrule
            \end{tabular}%
            \label{tab:baseline_method_info}
        }
    \end{table}

    \subsection{Parameter settings}

    This subsection primarily describes the parameter configurations for the proposed model and the associated experimental evaluations.
    For the hyperparameter settings of the proposed MPDFormer, the period resolutions $[64, 40, 112]$ and $[72, 56]$ for CDR and RDR datasets, respectively, are derived based on the spectrum offset of all ZigBee signals at 20dB.
    The quantity of encoding layers deployed for both two datasets is uniformly assigned a value of $5$.
    In the MPDFormer architecture, the number of attention heads is designated as $1$ for both the CDR and the RDR datasets.
    The parameters $\sigma=1.0$ and $\varsigma=0.1$ are calibrated to modulate the balance between inter-period and intra-period attention mechanisms.

    As shown in Table~\ref{tab:parameters_of_training_device}, the training and testing of the proposed model utilized an NVIDIA GeForce RTX 2080 Ti, which is equipped with 11 GB of GDDR6 memory.
    This graphics card, renowned for its superior processing power, significantly accelerates the training and testing phases of our deep learning models.
    In parallel, the deployment experiments in real-world scenarios were conducted using a NVIDIA Jetson Orin NX 16GB device.
    This device integrates an ARM Cortex-A78AE CPU and a NVIDIA Ampere architecture GPU, delivering 100 TOPS of processing capability which is essential for handling complex computations in dynamic environments.

    \begin{table}[h]
        \centering
        \caption{Comparative of NVIDIA GeForce RTX 2080 Ti and Jetson Orin NX}
        \label{tab:parameters_of_training_device}
        \resizebox{\columnwidth}{!}{
            \begin{tabular}{c|cc}
                \toprule
                \textbf{Attribute} & \textbf{GeForce RTX 2080 Ti}   & \textbf{Jetson Orin NX}         \\
                \midrule
                CPU                & Core i7-8700k                  & ARM Cortex-A78AE                \\
                GPU                & NVIDIA GeForce RTX 2080 Ti     & NVIDIA Ampere                   \\
                RAM                & 56 GB                          & 16 GB                           \\
                Power Usage        & 650W                           & 15W/20W/25W                     \\

                Purpose            & Training and testing of models & Real-world deployment of models \\
                \bottomrule
            \end{tabular}
        }
    \end{table}

    For the experimental settings, the PyTorch deep learning framework, in synergy with the CUDA acceleration library, was employed to leverage the computational prowess of the GeForce RTX 2080 Ti at training and testing process.
    To guarantee experimental reproducibility, the experimental parameter settings, including learning rate, batch size, optimizer, etc, are presented in Table~\ref{tab:experiment_settings}.

    \begin{table}[htbp]
        \small
        \caption{Experimental Settings for MPDFormer.}
        \centering
        \begin{tabular}{c|c}
            \toprule
            \textbf{Settings}     & \textbf{Value}           \\
            \midrule
            Scheduler             & Cosine Decay             \\
            Maximum Learning Rate & CDR: 0.0012, RDR: 0.0007 \\
            Optimizer             & AdamW                    \\
            Warmup Steps          & 4,080 (15 epochs)        \\
            Weight Decay          & 0.00001                  \\
            Batch Size            & 512                      \\
            Epochs                & 300                      \\
            \bottomrule
        \end{tabular}%
        \label{tab:experiment_settings}%
    \end{table}

    \subsection{Numerical results}

    \subsubsection{Comparison with SOTA methods}
\begin{figure}[h]
        \centering
        \begin{minipage}{0.47\textwidth}
            \centering
            \includegraphics[width=\linewidth]{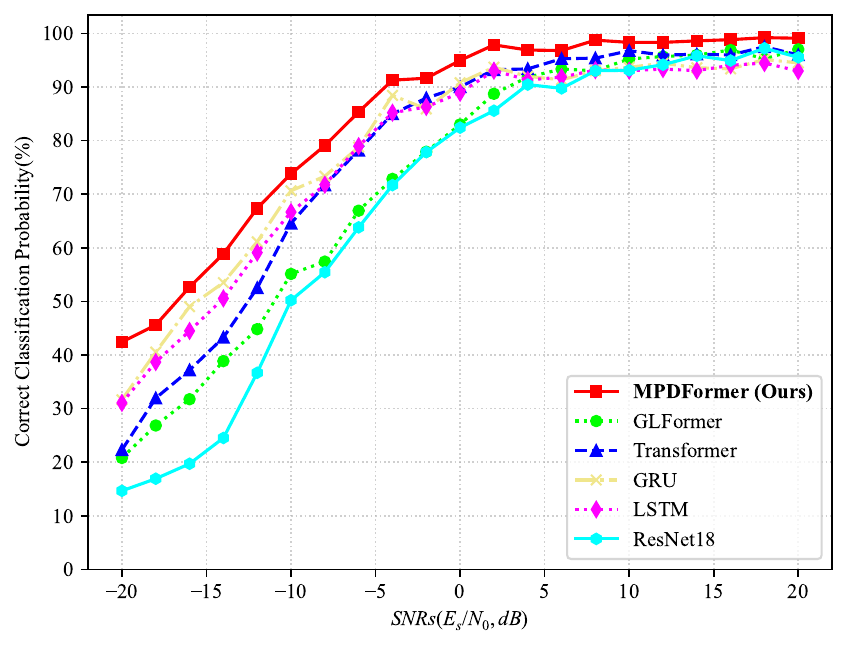}
            \caption{Comparison of the proposed MPDFormer against several baseline RFFI methods at RDR dataset.}
            \label{fig:performance_comparison_with_exist_RFFI_methods_at_DCR_dataset}
        \end{minipage}
        \hfill
        \begin{minipage}{0.47\textwidth}
            \centering
            \includegraphics[width=\linewidth]{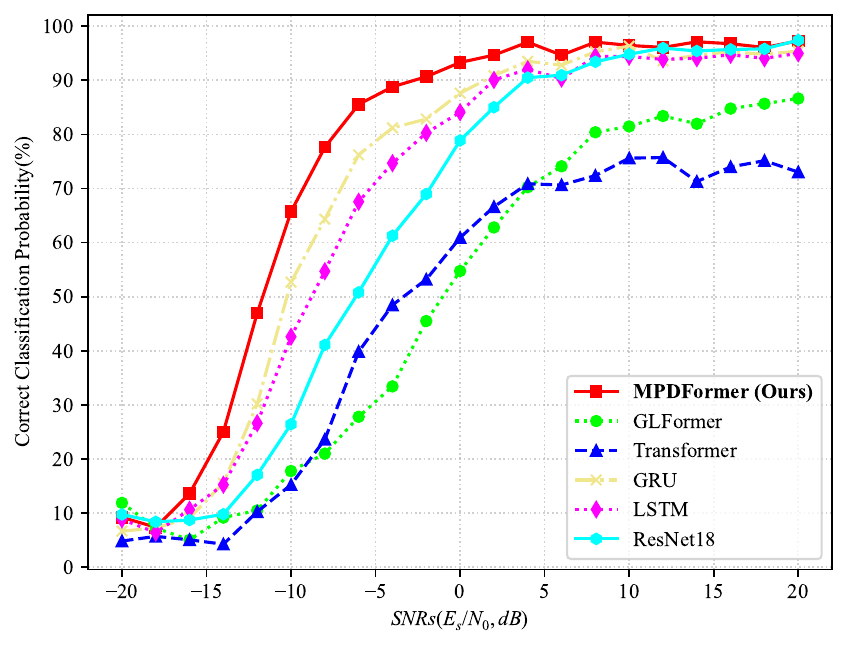}
            \caption{Comparison of the proposed MPDFormer against several baseline RFFI methods at CDR dataset.}
            \label{fig:performance_comparison_with_exist_RFFI_methods_at_DRR_dataset}
        \end{minipage}
    \end{figure}

    As is shown in Fig.~\ref{fig:performance_comparison_with_exist_RFFI_methods_at_DCR_dataset} and Fig.~\ref{fig:performance_comparison_with_exist_RFFI_methods_at_DRR_dataset}, the proposed MPDFormer exhibits the superior performance over other baseline methods at both CDR and RDR datasets.
    Specifically, the experimental results present that the proposed method outperforms these baseline methods across the SNR range from -20dB to 20dB at CDR dataset.
    At RDR dataset, MPDFormer demonstrates superior classification performance compared to the baseline method across a range from -16dB to 20dB, with particularly notable enhancements in classification efficacy between -14dB and 4dB.
    Moreover, our proposed method achieve more than 90\% accuracy when the SNR $\geq 0$dB.
    Therefore, the experiment results reveal that MPDFormer not only outperforms the baseline methods but also exhibits particularly pronounced efficacy under low SNR scenario.
    The results attribute the periodicity-dependency attention mechanism with period-delay correlation and temporal-delay correlation, which enhances the RFF features across multiple specific periods, and concurrently suppresses background noise and features of weak-periodicity or non-specified periodicity.

    \subsubsection{Ablation study of component analysis}

    (1) Ablation study of periodic embedding representation

    As is shown in Table~\ref{tab:spectrum_offset_under_single_period} and Table~\ref{tab:spectrum_offset_under_multiple_periods}, the ablation study investigates the impact of utilizing spectrum offset-based periodic embedding representation in the MPDFormer method on the RDR dataset under both single and multiple period conditions.

    In Table~\ref{tab:spectrum_offset_under_single_period}, which assesses the model's performance in a single period setting, it is suggested that utilizing PER significantly enhances the model's accuracy. With PER, accuracies are recorded at 73.8\% and 59.0\% for a single period of 72 and 56, respectively. In contrast, the configurations without PER show drastically reduced effectiveness, with accuracies plummeting to as low as 6.6\%, 8.7\%, 71.4\%, and 67.6\% for the period of 16, 32, 64, and 128, respectively. This indicates that near the 72 and 56 periods where the PER is employed, the performances at 64 and 128 periods are relatively superior, while the results for other periods demonstrate significant degradation. Therefore, these results highlight the pivotal role of PER in bolstering the model's capability to handle single-period spectrum more efficiently.

    Table~\ref{tab:spectrum_offset_under_multiple_periods} extends the analysis to multiple period settings. The results corroborate the advantageous impact of PER, with the model achieving a high accuracy of 74.8\% when PER is applied over multiple period resolutions [72, 56]. Without PER, the model’s performance varies more broadly across different combinations, reaching accuracies of 42.0\%, 72.9\%, 68.7\%, and a markedly lower 10.8\% for the broadest period combination [128, 256].
    These findings further illustrate that performance tends to degrade both at shorter and longer periods, with only those periods close to the PER settings achieving comparably high accuracy. This underscores the effectiveness and guiding role of PER in optimizing model performance.

    In conclusion, the ablation studies affirm that employing spectrum offset-based periodic embedding representation in the MPDFormer method enhances overall performance, with even greater benefits observed when extending to multiple period resolutions.

    \begin{table}[h]
        \centering
        \small
        \caption{The ablation study of periodic embedding representation within a single period at RDR dataset}
        \begin{tabular}{@{}c|cc|ccccc@{}}
            \toprule
            \multirow{2}{*}{Single Period} & \multicolumn{2}{c}{w/ PER} & \multicolumn{4}{c}{w/o PER} \\
            \cmidrule(l){2-8}
            & 72   & 56   & 16  & 32  & 64   & 128  & 256  \\
            \midrule
            Accuracy (\%) & \textbf{73.8} & 59.0 & 6.6 & 8.7 & 71.4 & 67.6 & 11.2 \\
            \bottomrule
        \end{tabular}
        \label{tab:spectrum_offset_under_single_period}%
    \end{table}

    \begin{table}[h]
        \centering
        \small
        \caption{The ablation study of periodic embedding representation within multiple periods at RDR dataset}
        \begin{tabular}{c|c|ccccc}
            \toprule
            \multirow{2}{*}{Multiple Periods} & \multicolumn{1}{c}{w/ PER} & \multicolumn{4}{c}{w/o PER} \\
            \cmidrule(l){2-6}
            & [72, 56] & [16, 32] & [32, 64] & [64, 128] & [128, 256] \\
            \midrule
            Accuracy (\%) & \textbf{74.8}     & 42.0     & 72.9     & 68.7      & 10.8       \\
            \bottomrule
        \end{tabular}
        \label{tab:spectrum_offset_under_multiple_periods}%
    \end{table}

    (2) Ablation study of the periodicity-dependency attention mechanism

    The ablation study presented in Table~\ref{tab:ablation_module} meticulously examines the influence of various modules within the MPDFormer on classification accuracy for both CDR and RDR datasets.
    This examination is pivotal in discerning the essential components that significantly contribute to the proposed model's performance.
    The study is structured around two principal components: intra-period and inter-period attention mechanisms.
    The results of these experimental investigations are summarized as follows.

    Initially, the baseline configuration, labeled MPDFormer$^1$ and characterized by the absence of two components, sets a foundational benchmark with classification accuracies of $67.5\%$ and $41.2\%$ at the CDR and RDR datasets, respectively.
    This baseline serves as a critical reference for evaluating the incremental improvements achieved in subsequent experimental configurations.
    The primary aim of the configurations MPDFormer$^2$ and MPDFormer$^3$ is to illustrate the distinct roles played by two attention mechanisms within the proposed model.
    MPDFormer$^2$ demonstrates a marked improvement in classification accuracy, achieving $77.0\%$ and $47.8\%$ for the CDR and RDR datasets, respectively.
    Building on the advancements made by MPDFormer$^2$, the MPDFormer$^3$ configuration further amplifies the model's classification capabilities, reaching an impressive accuracy of $84.1\%$ and $74.8\%$ for the CDR and RDR datasets, respectively.
    These results underscore the significance of the integrated attention mechanisms in enhancing the overall performance of the MPDFormer model.
    Therefore, these two modules play an integral role in the multi-periodicity dependency Transformer architecture, serving as indispensable RFF feature extraction modules within the MPDFormer model.

    \begin{table}[htbp]
        \centering
        \small
        \caption{Ablation study of the periodicity-dependency attention mechanism of MPDFormer. \XSolidBrush means: No intra-period or inter-period attention module exist in MPDFormer.}
        \resizebox{\columnwidth}{!}{
            \centering
            \begin{tabular}{c|ccccc}
                \toprule
                \textbf{Method} & \textbf{Intra-Period Attention} & \textbf{Inter-Period Attention} & \textbf{CDR(\%)} & \textbf{RDR(\%)} \\
                \midrule
                MPDFormer$^1$   & \XSolidBrush                    & \XSolidBrush                    & 67.5             & 41.2             \\
                MPDFormer$^2$   & \Checkmark                      & \XSolidBrush                    & 77.0             & 47.8             \\
                MPDFormer$^3$   & \Checkmark                      & \Checkmark                      & \textbf{84.1}    & \textbf{74.8}    \\
                \bottomrule
            \end{tabular}%
            \label{tab:ablation_module}%
        }
    \end{table}%

    \subsubsection{Sensitive analysis of hyperparameters}

    (1) Analysis of Top $k$ period resolutions

    We have conducted experiments to evaluate the impact of the period resolution $k$ on the classification performance of the MPDFormer method.
    The experimental results demonstrate that the MPDFormer achieves optimal classification performance on the CDR dataset with $k=3$ and on the RDR dataset with $k=2$, as shown in Table.~\ref{tab:comparison_with_top_k}.
    This finding is consistent with the spectral sparsity analysis conducted in subsection~\ref{subsec:spectrum-offset-&-multi-periodicity-processing}, reinforcing the theoretical basis for these specific values of $k$.
    From these results, it is presented that the performance differences across various $k$ values are not dramatically varied, suggesting a moderate sensitivity to this particular hyperparameter.
    Notably, $k=1$ also delivers commendable results, closely aligning with the best performance observed, indicating the effectiveness of the spectrum offset and its derived periods.
    Additionally, for values of $k$ greater than 3, while encompassing the characteristics of $k=2$ and $k=3$, the classification performance tends to decline due to the influence of higher frequency noise.
    Therefore, the choice of $k$ impacts the optimal performance of the MPDFormer, but overall, this hyperparameter is not sensitive.

    \begin{table}[htbp]
        \small
        \centering
        \caption{Comparison of various period resolutions $k$.}
        \begin{tabular}{c|ccccc}
            \toprule
            \multicolumn{1}{c|}{\multirow{2}[4]{*}{\textbf{Dataset}}} & \multicolumn{5}{c}{\textbf{Period Resolutions $k$ (\%)}} \\
            \cmidrule{2-6}    \multicolumn{1}{c|}{} & 1    & 2             & 3             & 4    & 5    \\
            \midrule
            CDR                                 & 82.9 & 82.9          & \textbf{84.1} & 83.2 & 83.3 \\
            RDR                                 & 73.8 & \textbf{74.8} & 73.4          & 74.4 & 74.4 \\
            \bottomrule
        \end{tabular}%
        \label{tab:comparison_with_top_k}%
    \end{table}%

    (2) Analysis of the number of encoding layer in MPDFormer

\begin{figure} [h]
        \begin{center}
            \includegraphics[width=0.65\textwidth]{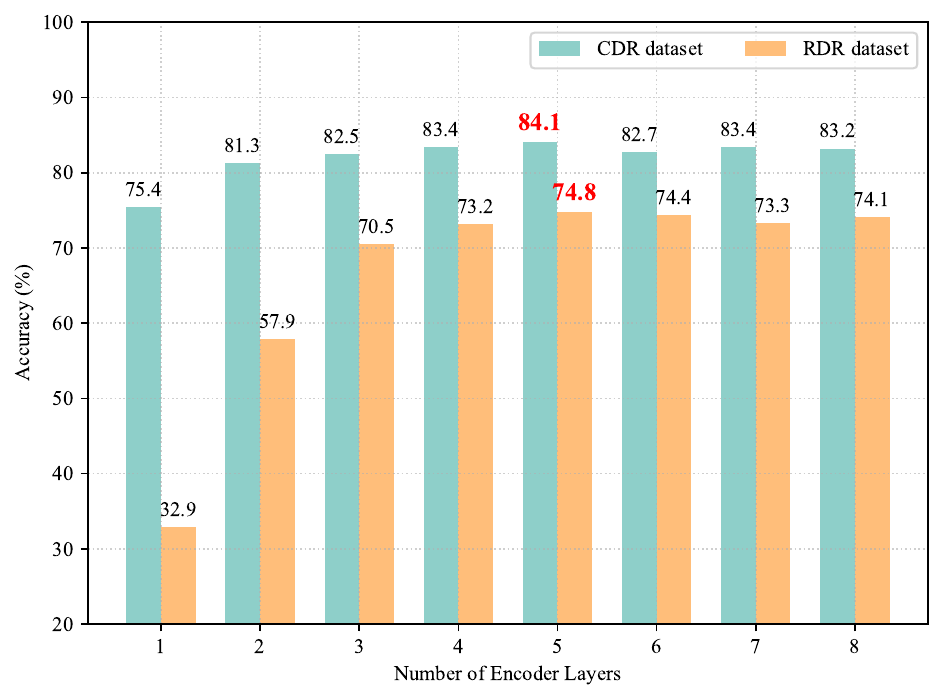}
        \end{center}
        \caption{Performance comparison of various numbers of layers in the MPDFormer encoder.}
        \label{fig:comparison_of_encoder_layers}
    \end{figure}

    As illustrated in Fig.~\ref{fig:comparison_of_encoder_layers}, the MPDFormer demonstrates a continuous improvement in classification performance as the number of encoding layers increases, with optimal performance observed at five layers.
    This finding indicates that increasing the number of layers initially enhances the model's ability to extract and utilize RFF features effectively.
    However, the performance peaks at five layers and slightly declines as additional layers are added, suggesting potential overfitting issues when the model becomes too complex.
    Despite the slight performance decrease beyond five layers, the drop is not severe, thanks to the regularization effects embedded within the AdamW optimizer used in the training process, which helps mitigate the overfitting effect.

    This pattern across CDR and RDR datasets suggests that the sensitivity of the MPDFormer model to the number of encoding layers is relatively mild, and the number of layers does not drastically affect the overall robustness of the model.
    To ensure optimal performance, selecting an appropriate number of encoder layers is crucial.
    The experimental results indicate that five layers provide a balanced approach for effective feature extraction without excessively complicating the model.
    Therefore, we recommend five encoding layers for achieving the best trade-off between complexity and performance in using the proposed MPDFormer.

    \begin{figure*}[t]
        \centering
        \begin{minipage}{0.32\textwidth}
            \centering
            \includegraphics[width=\linewidth]{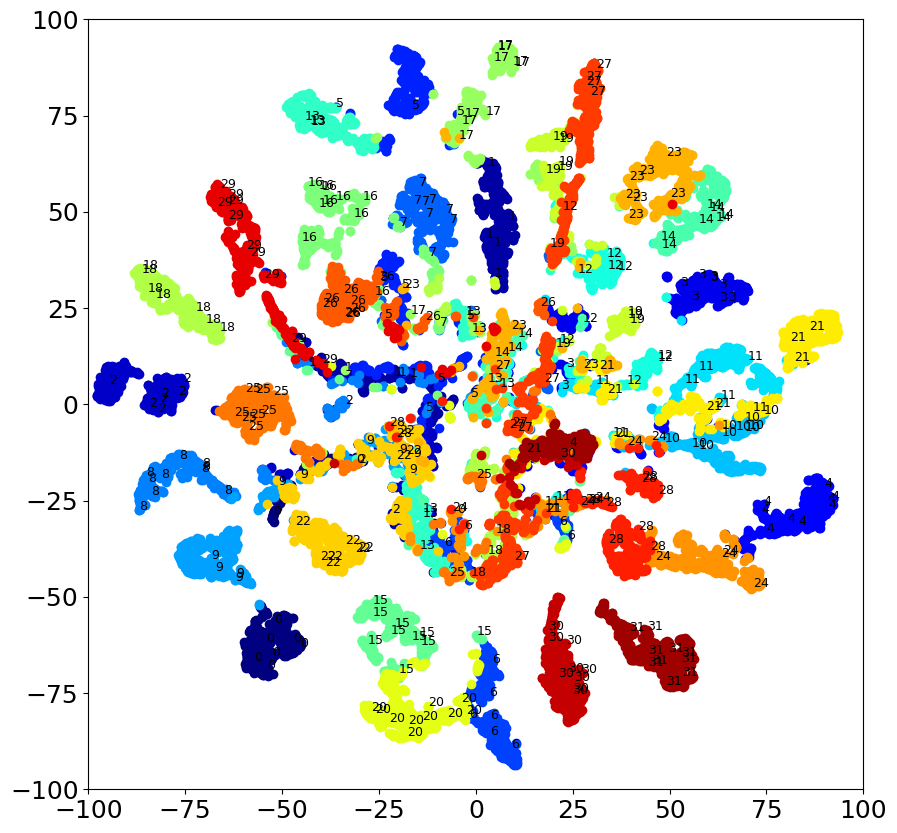}
            \subcaption{\textbf{MPDFormer (Ours)}}
            \label{fig:sne_MPDFormer_steady}
        \end{minipage}
        \hfill
        \begin{minipage}{0.32\textwidth}
            \centering
            \includegraphics[width=\linewidth]{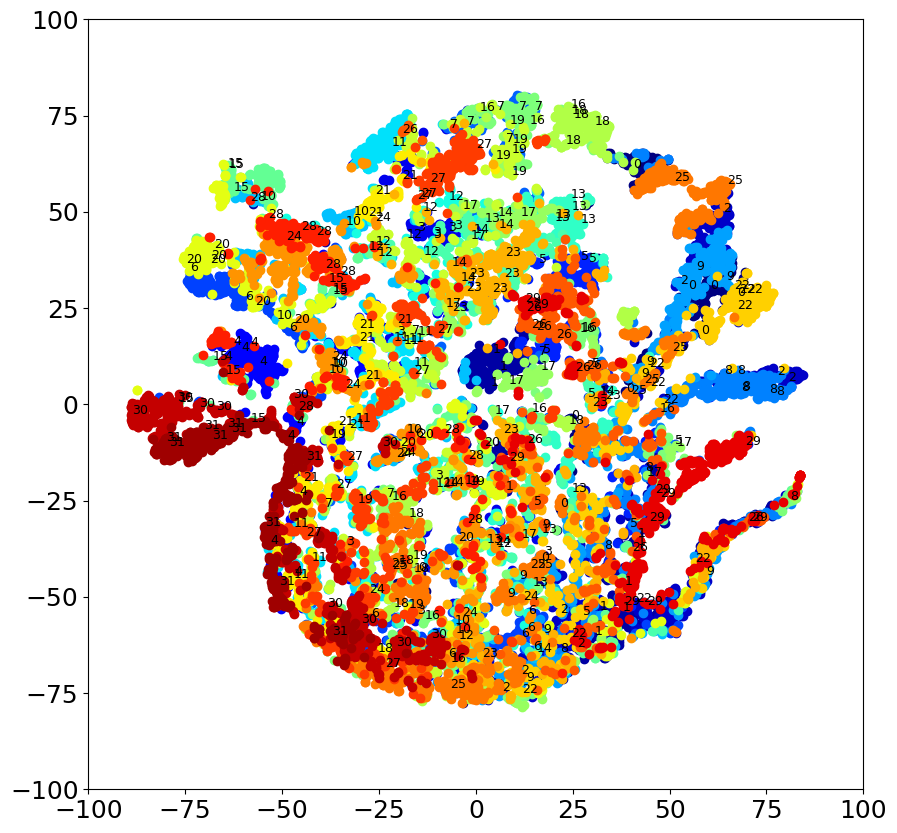}
            \subcaption{GLFormer}
            \label{fig:sne_GLFormer_steady}
        \end{minipage}
        \hfill
        \begin{minipage}{0.32\textwidth}
            \centering
            \includegraphics[width=\linewidth]{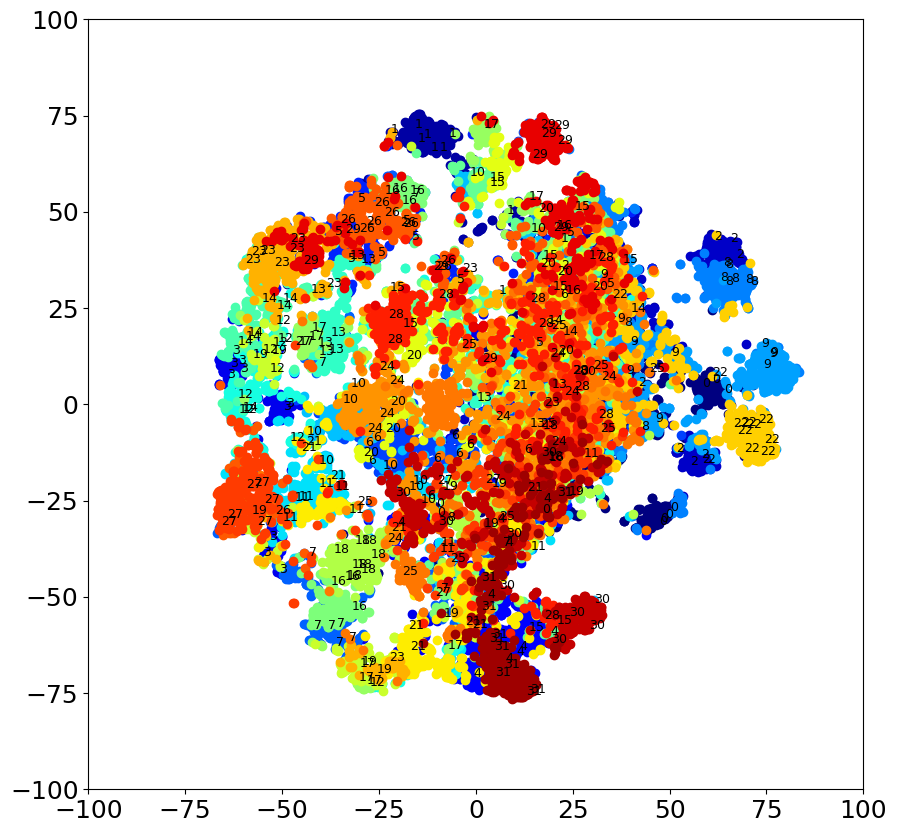}
            \subcaption{Transformer}
            \label{fig:sne_Transformer_steady}
        \end{minipage}

        \vspace{10pt}

        \begin{minipage}{0.32\textwidth}
            \centering
            \includegraphics[width=\linewidth]{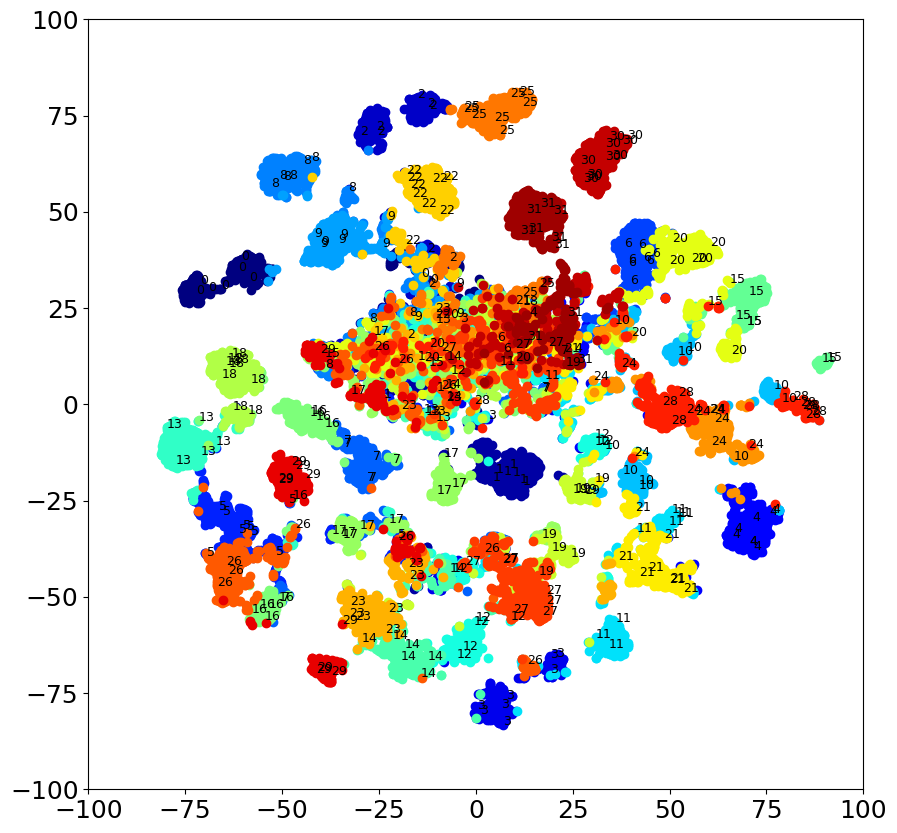}
            \subcaption{GRU}
            \label{fig:sne_GRU_steady}
        \end{minipage}
        \hfill
        \begin{minipage}{0.32\textwidth}
            \centering
            \includegraphics[width=\linewidth]{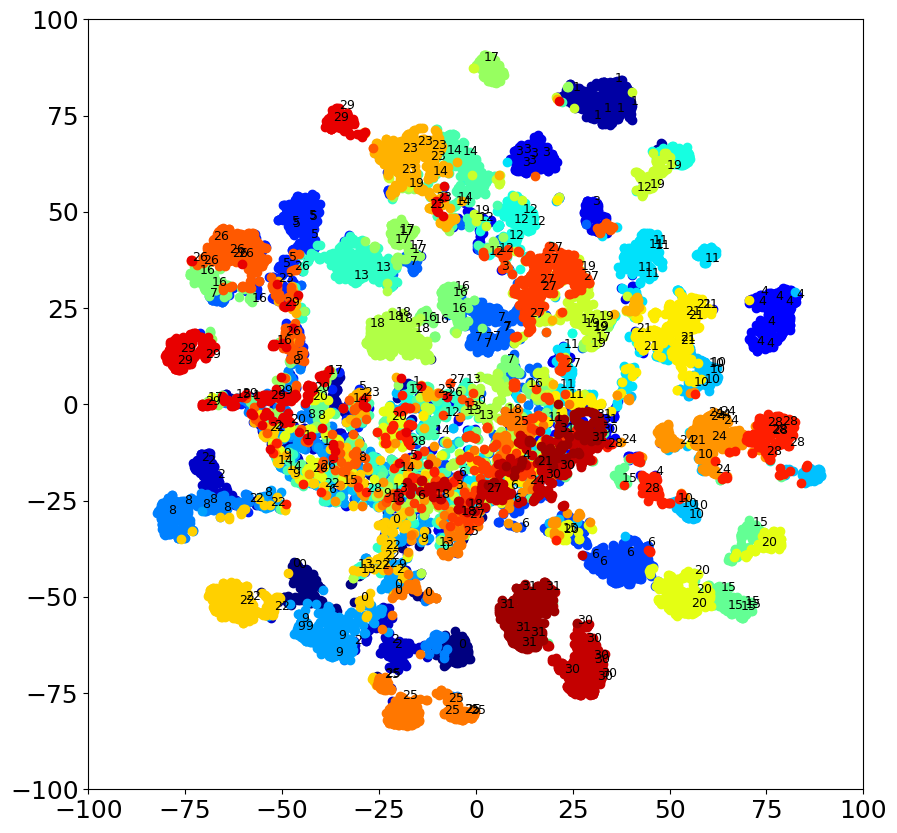}
            \subcaption{LSTM}
            \label{fig:sne_LSTM_steady}
        \end{minipage}
        \hfill
        \begin{minipage}{0.32\textwidth}
            \centering
            \includegraphics[width=\linewidth]{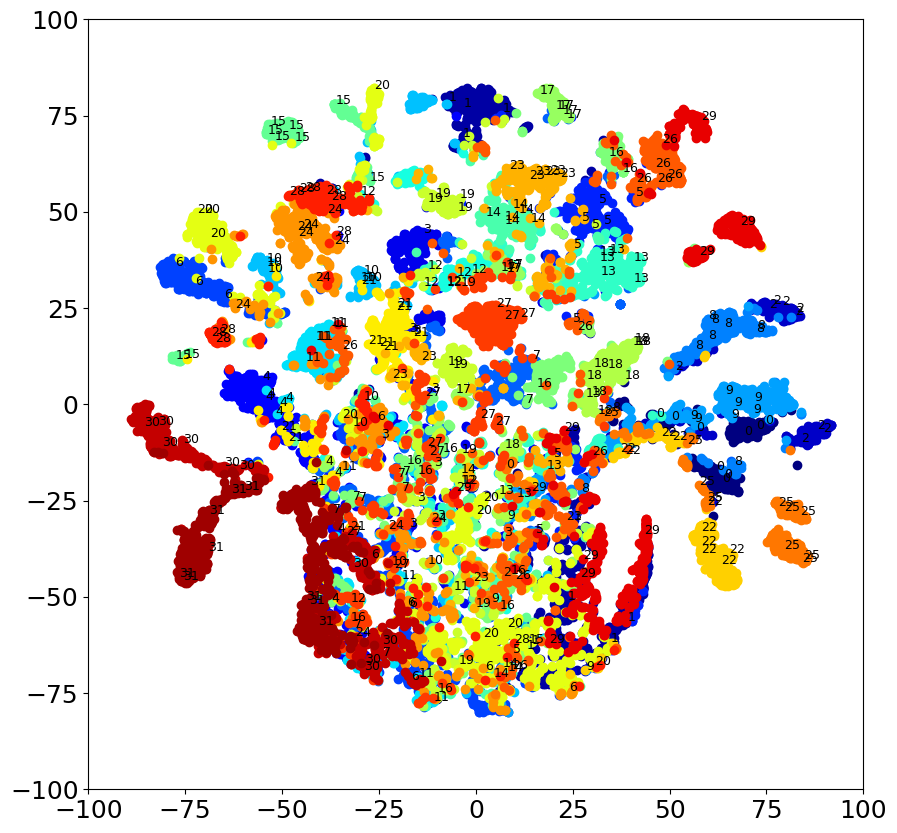}
            \subcaption{ResNet18}
            \label{fig:sne_ResNet18_steady}
        \end{minipage}

        \caption{T-SNE visualization of various RFFI methods at the SNR range of -20dB to 20dB within RDR dataset. The digital numbers (0-31) mean the label of 32 ZigBee devices.}
        \label{fig:t_sne_steady_among_all_methods}
    \end{figure*}

    (3) Analysis of the number of heads in MPDFormer

    Table~\ref{tab:comparison_of_number_of_heads} presents the experimental findings across different head counts in the MPDFormer applied to both CDR and RDR datasets.
    The experimental results indicate that increasing the number of heads beyond one does not yield improvement in classification accuracy.
    Indeed, the optimal performance was achieved consistently with a single head across both datasets.
    This suggests that the multi-head mechanism, while beneficial in typical Transformer applications for parallel processing of features, may not be ideally suited to our method which focuses on extracting and leveraging periodic embedding representation from RF signals.

    A plausible explanation for this observation is the fragmentation of the periodic embedding across multiple heads.
    Each head processes only a fraction of the embedding, thereby reducing the effective dimensionality available for capturing and correlating signal features.
    This diminished capacity likely disrupts the signal's periodic integrity, undermining the model's ability to establish meaningful contextual connections across and within signal periods.
    Therefore, we recommend restricting the number of heads to one or two to minimize the disruption of signal periodicity given these findings.

    Despite the potential for fragmentation and disruption of periodic signals caused by multiple heads, our model demonstrates good robustness to performance degradation due to this hyperparameter.
    This robustness suggests that the impact of increasing the number of heads on the MPDFormer's performance is not particularly sensitive, confirming that MPDFormer maintains effectiveness even when facing the challenges posed by multi-head configurations.

    \begin{table}[htbp]
        \small
        \centering
        \caption{Comparison of various numbers of heads at periodicity-dependency attention mechanism.}
        \begin{tabular}{c|ccccc}
            \toprule
            \multicolumn{1}{c|}{\multirow{2}[4]{*}{\textbf{Dataset}}} & \multicolumn{5}{c}{\textbf{Number of Heads (\%)}} \\
            \cmidrule{2-6}    \multicolumn{1}{c|}{} & 1             & 2    & 4    & 8    & 16   \\
            \midrule
            CDR                                 & \textbf{84.1} & 82.7 & 83.4 & 80.4 & 82.7 \\
            RDR                                 & \textbf{74.8} & 73.7 & 72.9 & 69.7 & 70.7 \\
            \bottomrule
        \end{tabular}%
        \label{tab:comparison_of_number_of_heads}%
    \end{table}%

    \subsubsection{T-SNE visualization}

    As is presented in Fig.~\ref{fig:t_sne_steady_among_all_methods}, we conducted a series of comparative experiments against baseline methods and visualized the performance of clustering and classifying wireless devices using the t-SNE technique at SNR range of -20dB to 20dB within the RDR dataset.
    The numbers in Fig.~\ref{fig:t_sne_steady_among_all_methods} represent the labels of 32 ZigBee devices.
    To ensure a fair comparison, all visualized results are constrained and presented within the coordinate system of [-100, 100].
    Compared to the baseline methods, our proposed method exhibits a more distinct and compact clustering performance shown in Fig.~\ref{fig:sne_MPDFormer_steady}, which suggests that the RFF features extracted by MPDFormer exhibit a more compact feature distribution of device categories within the same class and a more distinct feature separation between different classes.
    Although our method experiences some category confusion in the middle section due to noise interference, compared to the extensive category confusion observed in other baseline methods, our proposed method demonstrates superior intra-class clustering and inter-class classification performance.
    Therefore, in the domain of RFFI, the proposed MPDFormer exhibits robustness and a higher tolerance to noise interference, making it highly suitable for practical application due to its exceptional clustering and classification capabilities.

    \subsubsection{Confusion matrix of MPDFormer}
\begin{figure} [h]
        \begin{center}
            \includegraphics[width=0.8\textwidth]{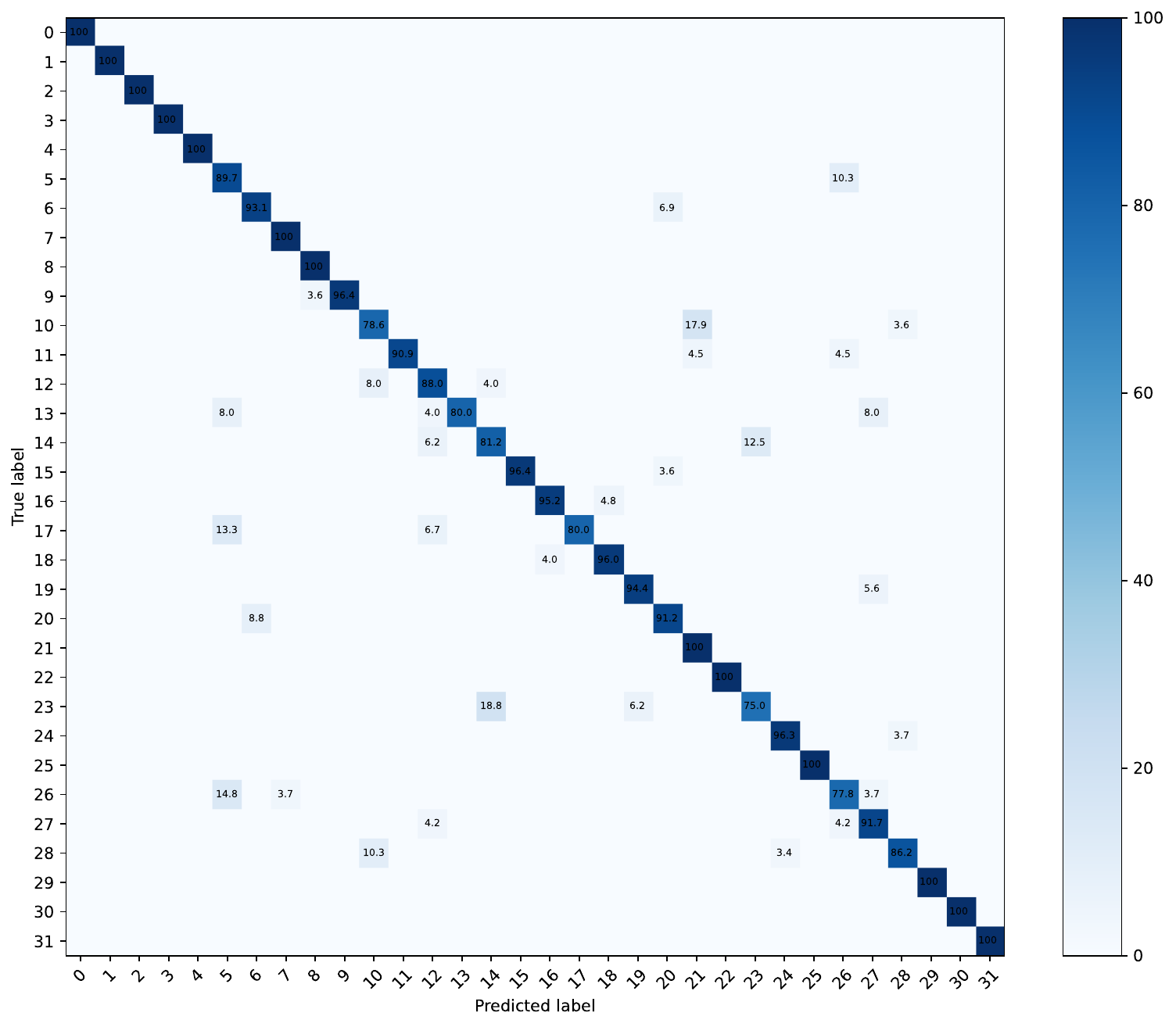}
        \end{center}
        \caption{The confusion matrix of the proposed MPDFormer with SNR=0dB at RDR dataset. Each cell in the matrix shows the percentage accuracy of the model in predicting the device label, where the row represents the true label and the column represents the predicted label.}
        \label{fig:confusion_matrix}
    \end{figure}

    This confusion matrix shown in Fig.~\ref{fig:confusion_matrix} evaluates the MPDFormer's efficacy in identifying ZigBee devices at SNR = 0dB on the RDR dataset.
    High values along the matrix's diagonal indicate accurate predictions across multiple classes (e.g., 100\% for labels 0, 1, 2, 3, 4, 7, 8, etc), showcasing the model's competence.
    However, non-diagonal entries highlight misclassifications, such as label 10 and 23 being mistaken for label 21 (17.2\% error) and label 14 (17.6\% error), respectively, suggesting difficulties in distinguishing similar signal patterns.

    The MPDFormer demonstrates substantial identification capabilities even in noisy conditions, yet the misclassifications indicate opportunities for refinement.
    Enhanced feature engineering and adjustments in training methods, such as applying class-weighting, may improve accuracy.
    Additionally, increasing data diversity through augmentation could fortify the model’s feature recognition, particularly for frequently misclassified classes, leading to more reliable performance.

    \subsubsection{Deployment experiment of MPDFormer}

    The MPDFormer model, as employed on the NVIDIA Jetson Orin NX, exhibits notable performance characteristics across different datasets shown in Table~\ref{tab:mpdformer_performance}.
    For the RDR dataset, which utilizes an input dimension of 2048 $\times$ 2 and a period resolution of 2, the model leverages 1,758,752 parameters and achieves a computational complexity of 54.7 MFLOPs. Importantly, it maintains an inference time of just 0.05 seconds.
    Conversely, the model in the CDR dataset, having a greater number of parameters at 3,817,536 due to a period resolution of 3, results in lower computational complexity (47.8 MFLOPs), but has an even longer inference time of 0.07 seconds.
    The primary reason for this lies in the presence of operations within the model that cannot be parallelized for acceleration, such as the fast Fourier transform in the periodicity-dependency attention mechanism.
    Nevertheless, both models operate at millisecond level for the inference time.

    Given the millisecond-level inference times, the proposed MPDFormer model demonstrates excellent real-time performance capabilities, making it highly suitable for deployment in actual communication scenarios and integration with the existing embedded system.
    This performance ensures that the model can be effectively implemented and utilized in real-world applications where timely data processing is critical.

    \begin{table}[ht]
        \centering
        \caption{The complexity and inference time of MPDFormer on NVIDIA Jetson Orin NX}
        \label{tab:mpdformer_performance}
        \resizebox{\columnwidth}{!}{
            \begin{tabular}{@{}c|ccccc@{}}
                \toprule
                \textbf{Dataset} & \textbf{Input Dimension} & \textbf{Period Resolution} & \textbf{Parameters} & \textbf{MFLOPs} & \textbf{Inference Time (s)} \\
                \midrule
                RDR              & 2048 $\times$ 2          & 2                          & 1,758,752           & 54.7            & 0.05                         \\
                CDR              & 1024 $\times$ 2          & 3                          & 3,817,536           & 47.8            & 0.07                         \\
                \bottomrule
            \end{tabular}
        }
    \end{table}

    \section{Conclusion}
    In this paper, we address the issue of weak energy and subtle differences in the RFF features of ZigBee signal which diminishes the capability of RFFI methods in feature representation.
    we propose a novel multi-periodicity dependency networks that leverages the power of Transformer-based architecture for extracting discriminative RFF features.
    By measuring the differentiation of RFF features from the spectrum offset in the frequency domain and the embedding periodic representation in the time domain, our proposed method can maximize the differentiation of RFF features in the RF signals from various wireless devices.
    The MPDFormer leverages the correlation of inter-period and intra-period RFF features to effectively manage periodicity-dependency features across both long-range and short-range contexts, thereby enabling a comprehensive method that not only robustly extracts distinct RFF characteristics of wireless devices but also effectively suppresses weak-periodicity component such as noise, facilitating the representation of multi-periodic RFF features.
    Experimental results demonstrate that the MPDFormer outperforms baseline methods in terms of classification performance and offers distinct inter-class and compact intra-class performance for wireless devices.
    The MPDFormer deployed on a NVIDIA Jetson Orin NX showcases the inference time of 0.07s, highlighting its potential for practical application in the real-world scenarios.

    In our future research, we aim to improve the performance of the MPDFormer in cross communication scenarios, specifically in cases where UAV link signals exhibit data distribution shifts.
    The focus will be on investigating domain adaptation strategies for RFFI in diverse wireless communication environments.
%
%    \section*{Acknowledgments}
%    This work is supported by the National Natural Science Foundation of China (grant numbers U20B2042, 62201010); and the R\&D Program of Beijing Municipal Education Commission (grant number KM202310009003).

    \bibliographystyle{elsarticle-num}
    \bibliography{MPDFormer_for_RFFI}

\end{document}